\documentstyle[12pt,preprint,epsf]{aastex}
\def\omegacen{$\omega$-Centauri~}
\def\msat{\mu}
\def\pc{\, {\rm pc} }
\def\kpc{\, {\rm kpc} }
\def\msun{M_\odot}
\def\lsun{L_\odot}
\def\kms{\, {\rm km \, s}^{-1} }
\def\BF{\mathbf}
\def\bey{\begin{eqnarray}}
\def\eey{\end{eqnarray}}
\def\beq{\begin{equation}}
\def\eeq{\end{equation}}

\def\edcomment#1{\iffalse\marginpar{\raggedright\sl#1\/}\else\relax\fi}
\reversemarginpar

\begin{document}
\title{Dynamical friction for dark halo satellites: 
effects of tidal massloss and growing host potential}
\author{HongSheng Zhao\footnote{UK PPARC Advanced Fellowship and Visiting Professorship at Beijing  Observatory}}
\affil{University of St. Andrews, Physics and Astronomy, KY16 9SS, Scotland\footnote{Present address (hz4@st-andrews.ac.uk)}}
\affil{Institute of Astronomy, Cambridge, CB30HA, England}
\begin{abstract}
Motivated by observations of inner halo satellite remnants like the Sgr and \omegacen, 
we develop fully analytical models to study the orbital decay and tidal massloss of satellites on
eccentric orbits in an isothermal potential of a host galaxy halo.
The orbital decay rate is often severely overestimated if applying the ChandraSekhar's 
formula without correcting for (a) the evaporation and tidal loss of the satellite and (b) the
contraction of satellite orbits due to 
adiabatic growth of the host galaxy potential over the Hubble time.  As a satellite 
migrates inwards, the increasing halo density affects the dynamical friction in two opposite ways: 
(1) it boosts the number of halo particles 
swept in the satellite's gravitational "wake", hence increasing the drag on the satellite, 
and (2) it boosts the tide which "peels off" the satellite, and reduces the amplitude of the wake. 
These competing processes can be modeled analytically for a satellite with the help of 
an empirical formula for the massloss history.  The analytical model agrees with 
more traditional numerical simulations of tidal massloss and dynamical friction.
Rapid massloss due to increasing tides at smaller and smaller radius 
makes it less likely for streams or remnants of infalling satellites
to intrude the inner halo (as the Sgr stream and \omegacen) 
than to stay in the outer halo (as the Magellanic stream), hence any intermediate-mass central black holes of the satellites are also likely "hung up" at 
large distances as well.  It is difficult for the satellites' black holes to come close enough to
merge into the supermassive black hole in the center of the host potential unless 
the satellites started with (i) pericenters much smaller than the typical distances to present-day
observed satellites and with (ii) central density much higher than
in the often seen finite density cores of observed satellites.

\end{abstract}
\section{Introduction}

Current theory of galaxy formation favors the idea that galaxies form hierarchically
by merging smaller lumps or satellites.  The gravity from a moving satellite pulls behind it
a "wake" of particles of the host galaxy (ChandraSekhar 1943, Mulder 1983).  
This dynamical friction dissipates the orbital energy of the satellites so that they sink deep into the host galaxy potential well, where they are
disintegrated and virialized via baryonic feedbacks and tidal stripping.  These processes might have
determined the density profile of virialized halo of the host galaxy (Syer \& White 1998, Dekel et al. 2003).

There are about 150 and 300 globular clusters, and a few dozen dwarf satellites
of mass $10^{6-9}\msun$ in the Milky Way and M31 respectively.
It is tempting to associate these
dwarfs satellites and globular clusters as the markers/remnants of past 
hierarchical merging events.
Indeed there are several examples of possible streams of remnants in the Milky Way 
(Lynden-Bell \& Lynden-Bell 1995) including the recently found Galactic ring or Carnis Major dwarf galaxy,
traced by a grouping of globular clusters (Martin et al. 2004).  
A giant stream is found in the Andromeda galaxy (McConnachie et al. 2003, Ferguson et al, 2002).
Among these the Sagittarius dwarf galaxy stream (Ibata et al. 1997) between radius 10-50 kpc 
from the Galactic center is perhaps the best example.  
It brings in at least 5 globular clusters to the inner halo, including M54, 
the 2nd most massive cluster of the Milky Way.   
A still mysterious object is \omegacen, with about $10^6\lsun$ at 5 kpc from the Galactic center
which has the morphology of a globular cluster, but has multiple epoches
of star formation and chemical enrichment (see Gnedin et al. 2002 and 
the \omegacen symposium).  Another example is the G1 cluster, 
a very massive globular-like object with about $10^6\lsun$ at about 40 kpc
from the M31 center (Meylan et al. 2001). A system (NGC1023-13) almost
identical to G1 is also found in the S0 galaxy NGC1023, at a
projected distance of about 40 kpc from the host galaxy (Larsen 2001).  
Freeman (1993) suggested that such systems are the remnants of nucleated dwarf satellites,
with their outer tenuous dark matter and stars being removed by galaxy tides.  
M31 also has an unusual collection of 
clusters as luminous as \omegacen within 5 kpc in projection from its double-peaked center.
Given the above evidences or signs for infalling objects in our galaxy and M31,
it is interesting to ask whether some of the inner globulars could have
also been the result of mergers.  Beyond the Local Group,
minor mergers are sometimes speculated as the mechanism to deliver massive
black holes and gas material into the nucleus of an AGN to account
for the directions of the jets and nuclear dusty disks (Kendall, Magorrian, \& Pringle 2003).  A very interesting
related issue is whether giant black holes acquire part of the mass
by merging the smaller black holes in the nuclei of infalling satellites.

For this paper we revisit the basic theoretical questions: what is 
the condition for a dwarf galaxy to decay into the inner halo? 
What are the possible outcomes of tidal stripping of a dwarf satellite? 
How often do we get a system like \omegacen or a naked black hole near the host center?
The answers to these questions will help us to test the validity of the theory of hierarchical
merger formation of galaxies.  
The key mechanism for satellites to enter the inner galaxy
is dynamical friction, where the gravity of the satellite creates a wake 
of overdensities in the particle distribution of the host galaxy,
which in turn drags the motion of the satellite with a force
proportional to $m(t)^2$, where $m(t)$ is the mass of the satellite.  
Another process is tidal disruption, where the object sheds mass with each 
pericenter passage, and the remnants are littered along the orbit of the satellite.  
The above two processes compete with and regulate each other:
orbital decay increases the tidal field, which reduces the mass of the satellite, 
hence slows down the orbital decay.  Some examples of these effects
have been shown in Zhao (2002) in the case of \omegacen.

The analytical formula of ChandraSekhar  is widely used for 
gaining insights on dynamical friction because of the time-consuming nature of
the more rigorous N-body numerical simulation approach.
It is a customary practice in previous works to model the orbital decay of 
a satellite as a point mass of a fixed mass.  However,  
{\it the fixed-mass approximation is invalid}
and could seriously overestimate dynamical friction because of neglecting 
massloss ${dm(t) \over dt}$.  
It is essential in calculations of satellite orbits to model 
the dynamical friction and massloss together since they regulate each other.

In the past the massloss and the orbital decay are often modeled in the {\it ab initio} fashion, 
resulting coupled non-linear equations without simple analytical solutions.  
In such models,  
the satellite mass is often modeled as a function of the satellite's tidal radius,
hence various factors come in, including the orbital position of the satellite,
the density profile of the satellite
(e.g., Jiang \& Binney 2000, Zhao 2002, Mouri \& Taniguchi 2003, Kendall et al. 2003).
However, these complications are not always necessary since the massloss history is 
rather similar in simulations with very different initial conditions
(the mass is generally a stair-case like a function of the time), so 
could be parametrized in an empirical fashion, by-passing the uncertain assumptions of the satellite
initial profile.  This could be useful for exploring a large parameter space of the satellite
initial condition.

Another invalid approximation but common practice is to use a static potential for the host galaxy.
This again is unphysical since galaxy halos do grow in hierarchical formation scenario
partly because of galaxy merging, and partly because of the adiabatic contraction
of the baryonic disk and bulge; galaxy rotation curves $V_{\rm cir}(r,t)$ 
can change by significantly before and after the formation of baryonic disks and bulges in mass models for the 
Milky Way and M31 (Klypin, Zhao \& Somerville 2002) and in generic CDM simulations with baryons (e.g. Wright 2003).
The growing gravitational force tends to restrain the radial
excursions of the satellite while preserving the angular momentum.  
The dynamical friction or drag force 
is also proportional to the growing density $\rho(r,t)$ of ambient stars and dark particles
in the host galaxy. 

In fact, it is conceptually simple to incorporate massloss and growing potential 
while keeping the problem 
analytically tractable: the deceleration $-{dv \over dt}$ is simply proportional to the 
satellite mass $m(t)$ and the ambient density $\rho(r,t)$ of the host galaxy at the time $t$.  
Without massloss ChandraSekhar's formula would predict very efficient braking
of the orbits, enough to make a high-mass satellite of $\sim 10^{10}\msun$
(the mass of the LMC or M33 sized object) to decay from a circular
orbit at $\sim 100$kpc to the very center of 
a high brightness galaxy in a Hubble time, delivering remnants into the inner galaxy.
Here we study the effect of massloss and growing potential on the 
result of orbital decay, and the distribution of remnants.  
We present fully analytical results for calculating the decay rate for satellites
on eccentric orbits in a scale-free growing isothermal potential.

The structure of the paper is following: S2 gives the analytical formulation
of the problem, S3 presents results of application to globulars and dwarf satellites, 
S4 studies the relation between massloss and the satellite density profile.
S5 discusses the progenitors of Sgr and \omegacen, and S6 summarizes.

\section{Analytical orbital decay model for a shrinking satellite in a growing host}

Consider a satellite moving with a velocity ${\BF v_s}$ on a rosette-like orbit 
in a spherical host galaxy potential $\phi(r,t)$ with a rotation curve $V_{\rm cir}$.  
Assume it has an initial mass $m_i$, 
the orbital decay from an initial time $t=t_i$ to the present day $t=t_0$
can be modeled using ChandraSekhar's dynamical friction formula (Binney \& Tremaine 1987)
\beq\label{dv}
{d{\BF v_s} \over dt} = - {{\BF v_s} \over t_{\rm frc}} - {V_{\rm cir}^2 \over r^2}{\BF r} 
\eeq
where $t_{\rm frc}$ is the instantaneous dynamical friction time.
Manipulating the equation, we find 
the (specific) angular momentum $j(t)$ of the satellite decays as
\beq
{d {\BF j}(t) \over dt} = {\BF r} \times {d{\BF v_s} \over dt} = - {{\BF j}(t) \over t_{\rm frc}} , 
\qquad {\BF j}(t)= {\BF r}(t) \times {\BF v_s}(t).
\eeq
Here the dynamical friction time is given by
\beq
t_{\rm frc}^{-1} = 
\left[4\pi G\rho(r,t)\right] {G m(t) \over V_{\rm cir}^3} \xi(u), \qquad u(t) \equiv \left({|{\BF v_s}| \over V_{\rm cir}}\right)
\eeq
where $m(t)$ is the mass of the satellite, 
$\rho(r,t)$ is the density of host galaxy at the orbital radius $r(t)$,
and the effect of the satellite speed $|{\BF v_s}|$ is contained in 
$\xi(u)$, which is some dimensionless function 
of the rescaled satellite speed $u$.

Here we adopt 
a Singular Isothermal Spherical (SIS) host galaxy model with a rotation curve generally growing with time
$V_{\rm cir}(t)$; it is normalized by the present time value $V_{\rm cir}(t_0)=V_0$.  
In this model the potential and density are given by
\beq
\phi(r,t)=V^2_{\rm cir}(t) \ln r,\qquad
\rho(r,t) = {V^2_{\rm cir}(t) \over 4 \pi G r^2}.
\eeq
Since the stars and dark matter velocity distribution in the host galaxy 
is an isotropic Gaussian, the dimensionless velocity function
\beq
\xi(u) = u^{-3} \left[{\rm erf}(u)-{2u \over \sqrt{\pi}} \exp\left(-u^2\right)\right] \ln\Lambda
\approx {\ln\Lambda \over {4 \over 3}+u^3} ,
\eeq
where the dimensionless Coulomb logarithm $\ln \Lambda$ is typically between unity and ten.
It is easy to verify that our {\it simple approximation} for $\xi(u)$ works to 
10\% accuracy for $0 \le u=\left({|{\BF v_s}| \over V_{\rm cir}}\right)<\infty$, which is accurate enough in practice 
because real velocity distributions are often slightly non-Gaussian anyway.
Note for very low speed ${\xi(0) \over \ln\Lambda} ={4 \over 3\sqrt{\pi}} \approx {3 \over 4}$ 
and for circular motion ${\xi(1) \over \ln\Lambda} ={\rm erf}(1)-{2 \over e\sqrt{\pi}} 
\approx {3 \over 7}$.

Combining the above equations we find the decay rate for (the amplitude of) 
the angular momentum is given by
\footnote{the dynamical friction time $t_{\rm frc}$ is now expressed in terms of 
the angular momentum $j(t)$ instead of $r(t)$ with the help of 
substitutions $r(t) \rightarrow j(t)/[|{\BF v_s}(t)|\cos\alpha(t)]$
and $|{\BF v_s}(t)| \rightarrow u(t) V_{cir}$.}
\beq
{d j(t) \over dt} = - {j(t) \over t_{\rm frc}}, \qquad  
{1 \over t_{\rm frc}} = { G \msat(t) V_{\rm cir}(t) \over j(t)^2},
\eeq
where we define an effective bound mass of the satellite at time $t$ 
\beq
\msat(t) \equiv m(t) \beta(t), \qquad
\beta(t) = \xi(u) u^2 \cos^2(\alpha),
\eeq 
where $u(t)={|{\BF v_s}(t)| \over V_{\rm cir}(t)}$ and $\alpha(t)$ are the rescaled speed and the pitch angle of the orbit
at time $t$.   
As we can see, the effective mass $\msat(t)$ lumps together several time-varying 
factors.  It is proportional to the satellite mass $m(t)$ by
a dimensionless factor $\beta(t)$ which incorporates the varying efficiencies of dynamical friction
in an orbital epicycle.  Note $\beta(t)$ is an oscillating function of time of order unity, which we will come to in section 2.2.
Introducing this
{\it effective} mass simplifies the physics details because 
the braking rate of specific angular momentum $-dj(t)/dt$ is now simply proportional to $\msat(t)$.

Integrating over time, we find the change in the specific angular momentum between
time $t_i$ to $t_0$ is given by 
\beq
-\Delta j^2 =  j^2(t_i) - j^2(t_0) = 2  G \mu_i V_0 \tau, 
\qquad \tau \equiv (t_i-t_0) \gamma \equiv \int_{t_i}^{t_0}  
{\msat(t) V_{\rm cir}(t) \over \mu_i V_0} dt,
\eeq
where we define $\gamma$ as the reduction factor from fixed-mass satellite model, and define 
$\tau$ as the effective duration of dynamical friction,
$\mu_i$ and $V_0$ are the effective mass at initial time $t_i$ 
and the rotation curve at the present $t_0$.  
The above formula allows us to calculate
the evolution of the specific angular momentum $j(t)$ of a generally
{\it eccentric satellite orbit}.

It is interesting to ask
what the necessary condition is for an outer halo satellite to reach the inner galaxy
or even the center.  These satellites will be destroyed with their remnants 
in the inner galaxy, e.g., by the disk and/or bulge shocking.  
To facilitate the comparison with observations,
it is sometimes better to define an instantaneous orbital size
\footnote{We can make this definition irrespective of the eccentricity of the orbit, which we does not
enter our calculation explicitly.  Roughly speaking the orbital size $S(t)$ is the geometrical mean of
the pericenter radius and apocenter radius in a static potential.} 
\beq
S(t) \equiv {j(t) \over V_0}
\eeq
at time $t$.  Let the present orbital size 
$S_0={j_0 \over V_0} \le R_{\rm disk}\sim 15$kpc,
and define $j_{\rm disk}=R_{\rm disk}\times V_0$ as the critical angular momentum to reach
within the disk at the present time.  Manipulating eq. (8)
we find the initial angular momentum $j_i$ or orbital distance $S_i={j_i \over V_0}$ must satisfy
\beq
S_i =
\left[ S_0^2 + {2G \mu_i \tau \over V_0} \right]^{1 \over 2}, \qquad 0\le S_0 \le R_{\rm disk}, 
\eeq
where $\tau$ is given in eq. (8).
In  other words, satellites initially on orbits larger than $S_i$ would not be able to deliver
a globular or dwarf galaxy to the inner galaxy irrespective of mass loss assumed.

\subsection{A geometrical interpretation}

Eq. (8) is our main analytical result.  There is also an interesting
geometrical interpretation to it.  Consider, for example, a satellite of mass $m(t)$ on 
a {\it circular orbit} in a static SIS potential,  we have
\beq
\ln \Lambda \approx 2.5, \qquad \beta(t) = \xi(1) \times 1^2=0.42 \ln \Lambda \approx 1,
\qquad \msat(t)=m(t)\beta(t) \approx m(t),
\eeq 
where we took a typical value for the Coulomb logarithm (e.g., Penarrubia, Kroupa \& Boily 2002).
We can rewrite the angular momentum equation (eq. 8) as
\beq
\left| \pi \Delta S^2 \right| \approx  
{2\pi G \left<m\right> \over V_0^2} \left(  V_0 \Delta t\right), \qquad
\left<m\right> = {\int_{t_i}^{t_0} m(t) dt \over t_0-t_i} 
\eeq
where the l.h.s. is the area swepted by the decaying orbit, 
and in the r.h.s.  
$2\pi G \left<m\right> V_0^{-2}$ is the "circumference of influence" of the satellite,
and the $V_0\Delta t=V_0(t_0-t_i)$ is the length of the satellite's orbital path,
and the multiplication of the two is the area swept by the circumference of influence
of the satellite.  The above equation implies that the two areas are comparable.
Interestingly the decay of the orbits depends on the satellite mass  
through the approximation $\Delta S^2 \propto \left<m\right> \Delta t $,
i.e., it is the average mass of the satellite that determines the rate
of orbital decay.   Finally note eq.(8) applies to the evolution of specific angular momentum 
of an eccentric orbit in a time-varying potential. 
A complete description of the orbits should also include the evolution of the orbital energy,
which unfortunately is more complex analytically, and is not studied in detail here.
It is not essential for our conclusion, but is perhaps convenient to assume
efficient dynamical decay of the orbital energy 
during the pericentric passages hence the radial motion is damped and 
orbit circularizes at the end.  Adiabatic growth of the potential also tends to circularize
the orbits.

\subsection{Modeling the growing host potential} 

Classical singular isothermal model assumes a fixed potential or a time-invariant rotation curve.
In reality galaxies grow substantially from redshift of a few to now.  
The growth of the dark halo is scale-free in the hierarchical scenario,  
so it is plausible to approximate the growth of the rotation curve
as a power-law in time as follows,
\beq
V_{\rm cir}(t)  = V_0 \left({t \over t_0}\right)^{p},\qquad 0<t<t_0,
\eeq
where the rotation speed at some earlier time $0<t<t_0$ is generally smaller than 
the present day ($t=t_0=14$Gyr) value $V_0$.    
At the present epoch $V_{\rm cir}(t_0)=V_0=200\kms$ is appropriate for the Milky Way.
A few sample evolution models are shown in Fig.1a.  A value of ${1 \over 9} \le p \le {1 \over 3}$ implies
an evolution of 10\%-25\% in rotation speed from redshift one to now, which seems 
reasonable in models of the Milky Way before and after the formation of 
the disk and the bulge (Klypin, Zhao, Somerville 2001; Wright 2003).

\subsection{Modeling the shrinking satellite}

Previously the tidal stripping of the satellite is either modeled
by N-body simulations (e.g., Tsuchiya et al. 2003, 2004, Johnston et al. 1999)
or semi-analytically using tidal radius criteria (e.g., 
Zhao 2002, Kendall, Magorrian \& Pringle 2003). 
\footnote{Recently the latter approach has also been extended to model the orbital decay and
evaporation-induced massloss in the dense star cluster near the Galactic center
(Mouri \& Taniguchi 2003, McMillan \& Portegies-Zwart 2003). }
In both approaches one needs to assume
a rigorous description of the mass profile of the progenitor satellite.  
It is unclear whether this approach can model efficiently
the realistic large scatter in the circular velocity curves of observed dwarf galaxies.

Here we take a very different approach.
We model the mass of the progenitor galaxy as a simple function of 
time with two free parameters.  We do not need an explicit prescription
of the satellite density profile.  Motivated by the massloss history
typically seen in N-body simulations, 
the satellite mass is modeled to decay in a rate between exponential and linear massloss.
To simplify the calculations, we lump together several uncertain factors,
and simply assume the {\it effective} mass of the satellite is shed following the following recipe,
\bey
\msat(t) = \left<\beta\right>_im_i \left\{1 - \left[1-\left({\left<\beta\right>_0m_0\over \left<\beta\right>_im_i}\right)^{1 \over n}\right]\hat{t} \right\}^{n} & , & \hat{t}={t-t_i\over t_0-t_i}, \\\nonumber
m(t) = m_i \left\{1 - \left[1-\left({m_0\over m_i}\right)^{1 \over n}\right]\hat{t}_N \right\}^{n} 
& , &  \hat{t}_N=\hat{t}-{\sin 2N\pi \hat{t} \over 2N\pi}\approx \hat{t},
\eey
where 
$m_0$ and $m_i$ are the mass at the present $t=t_0=14$Gyrs and 
when the satellite falls in $t=t_i$, and 
$0 \le \hat{t} \le 1$ is the rescaled dimensionless time, and $N\sim 1-100$ 
is the total number of pericentric passages between time $t_0$ and $t_i$.
The parameters $\left<\beta\right>_i$ and  $\left<\beta\right>_0$ are the
initial and final effective mass convertion factors, which we will come to.

The parameter $n$ determines the profile of the massloss history.  It is tunable
with the $n=1$ model having a constant rate of massloss,
and the $n \rightarrow \infty$ having an exponential massloss;
and $n=-1$ is a roughly power-law decay.
In simulations we typically see somewhere in between these cases:
$0<n \le 1$ for Plummer or King model satellites with 
a sharp fall off after an initially linear massloss (e.g., Penarrubia et al. 2002); for isothermal 
satellite models, the massloss is close to linear $n=1$ (Zhao 2002).  
Note for a completely disrupted satellite $m_i \gg m_0 \sim 0$.   
In N-body simulations satellite losses mass mainly in bursts near pericentric passages,
so the mass is a descending staircase like function of time.  
This is simulated fairly well by our formula with $\hat{t}_N$ in $m(t)$:  
a few example massloss histories are shown in Fig.1a where
we assume $N=10$ pericentric passages from $t_i=4$Gyr to now $t_0=14$Gyr. 

\subsection{Modeling the effective mass conversion factor $\left<\beta\right>$}

The conversion factor between the effective mass $\mu(t)$ and the actual 
mass $m(t)$ of the satellite is given by 
\beq
\beta(t) = {\mu(t) \over m(t)} = \xi(u) \left(u\cos\alpha\right)^2,
\eeq
where $u$ is the rescaled satellite velocity at time $t$ and $\alpha(t)$ is the pitch angle.
To compute the average value of $\beta(t)$ 
over one epicycle, let us first define
$C$ as the ratio of the apocenter $r_a$ to pericenter $r_p$, 
or the ratio of pericenter velocity $u_aV_{\rm cir}$ and $u_pV_{\rm cir}$,
where 
\beq
u_p V_{\rm cir} = { j \over r_p }, \qquad C \equiv {r_a \over r_p} ={u_p \over u_a}.
\eeq
According energy conservation we have
\beq
{E \over V_{\rm cir}^2 } = 
{u_p^2 \over 2} + \ln r_p = {u_a^2 \over 2} + \ln r_a,
\eeq
where $E$ is the energy at pericenter and apocenter.  
It is easy to show that $r_p$ and $u_p$ are determined by the angular momentum $j$
and the circular velocity $V_{\rm cir}$ as follows,
\beq
r_p = j V_{\rm cir}^{-1} u_p^{-1}, \qquad u_p = \sqrt{2\ln C  \over 1- C^{-2} }.
\eeq
Now if we make the approximations that
\beq
{\mu (t) \over m(t) } \approx
\left<\beta(t)\right> \approx \xi(\left<u\right>) \left<u^2\cos^2\alpha\right>
\eeq
and the further approximations
\beq
\left<u^2\cos\alpha^2\right> \approx u_p u_a
\qquad \left<u\right> \approx 
\left({u_p^{2k}+u_a^{2k} \over 2}\right), \qquad k={9 \over 11}
\eeq
then we have
\beq
{\mu (t) \over m(t) } \approx \left<\beta(t)\right> 
\approx \xi\left(\left<u\right>\right) 
\cdot \left({2\ln C \over C-C^{-1}}\right), 
\qquad \left<u\right>={C^k+C^{-k} \over 2}
\left({2\ln C\over C-C^{-1}}\right)^{k}, \qquad k={9 \over 11}
\eeq
Eq. (21) is then checked with the time averages computed by 
direct numerical integration of the orbit from pericenter to apocenter, 
and is found to be 
accurate to 2\% for $1 \le C \le 100$ 
(A simpler expression with $k=1$ gives somewhat poorer accuracy).  
While we negelct variation
of $\beta$ within one epicycle,
there can still be secular evolution because the orbital energy is lost more efficiently at pericenter, hence the orbit tends to circularize, 
which increases the effective mass conversion factor $\left<\beta\right>$.
Generally speaking the factors 
$\left<\beta\right>$, $u_p$ and $u_a$ etc. are functions of $C$, which can be
read out from the curves in Fig.1b.  For those Milky Way satellites and globular
clusters which we have good proper motions, the C-values are typically between 1 and 10
(Dinescu et al. 1999), hence 
$\left<\beta\right>$ varies only by a factor of at most three for these satellites.
So the uncertainty of their effective masses $\mu =\left<\beta\right> m$ 
is still largely from errors of their observed mass-to-light ratios.

The above conversion factor 
should be a good approximation for eccentric satellite orbits with apo-to-peri ratio of $C$ 
in a flat rotation curve potential.  For example, for an apo-to-peri ratio $C={r_a \over r_p}=5$,
we have a rescaled pericenter speed $u_p \approx \sqrt{3}$, and an average mass conversion factor
$\left<\beta\right>  \approx 0.5 \times \left(0.42\ln \Lambda\right)$.  Assume $\ln \Lambda=2.5$
we have $\mu(t) \approx 0.5m(t)$.  In comparison $\mu(t) \approx m(t)$ for circular orbits.
So for the same angular momentum, our model suggests that 
eccentric orbits have smaller effective mass of the satellite,
hence slower evolution of the orbital angular momentum.\footnote{However, the circularization of orbit tends to enhance dynamical friction.}
It is interesting that all three factors: massloss, growing potential, and orbital eccentricity
all work in the same sense of reducing the evolution of orbital angular momentum.

For a rough estimation of the reduction of dynamical friction due to 
satellite massloss and host growth, let's 
consider a {\it Gedanken} experiment where a satellite 
enters a growing host galaxy at time $t_i=0$ 
right after the big bang and is completely dissolved ($\mu_0=0$) by the time $t_0$.
Neglecting the sinusoidal component by letting $N\rightarrow \infty$,
the reduction factor $\gamma$ is computed by substituting eqs. (13-14) into eq. (8).
We find $\gamma={p!n!\over (p+n+1)!} = {1 \over 6}$ if the satellite 
losses mass linearly with time ($n=1$) 
in a linearly growing halo ($p=1$). 
This estimate is perhaps to the extreme.  
In reality the formation and mergers of the satellites are probably 
over an extended period of the Hubble time, perhaps 
starting around redshift of 1.5 ($t_i=4$Gyr), ending around now ($t_0=14$Gyr).
Hereafter we consider mostly models with 
$t_i=4$Gyr, and $t_0=14$Gyr, $0 \le p \le {1 \over 3}$ 
and $\mu_0 \ll \mu_i$.

\section{Results of application to globulars and dwarf satellites}

Section 2 gives the formalism to predict analytically
the evolution of the angular momentum of the satellite 
for a satellite with any massloss history on an eccentric orbit
around a time-varying potential. 

Clearly not all satellites could reach the galactic center as a globular or a naked massive
black hole.  Dynamical friction is basicly turned off if the satellite bound mass drops 
below $10^9\msun$ before reaching the inner galaxy.
To reach the inner, say, 15 kpc of the host galaxy, which is roughly 
the truncation radius of the outer disk of the Milky Way,
a satellite must have presently a specific angular momentum 
\beq
0 \le j_0 \le j_{\rm disk} \equiv R_{\rm disk}V_0=15{\rm kpc}\times 200 \kms.
\eeq
In comparison the present specific angular momentum of some of the known
satellites are given in Table 1.  The Magellanic stream and Ursa Minor have  
$j_0 \sim 15000\kms$kpc much larger than $j_{\rm disk}$.

\begin{deluxetable}{llll}
\tablecaption{Orbital size of known satellites of the Local Group \label{tbl-1}}
\tablehead{ \colhead{Object} & \colhead{Spec. Ang. Mom.} & \colhead{Orbital size} &\colhead{Ref}}
\startdata
\colhead{}&\colhead{$j_0={\BF r} {\rm kpc} \times {\BF v} \kms$} & \colhead{$S_0 \equiv {j_0 \over 200}$kpc}&\colhead{}\\
\tableline
{\rm \omegacen}		& $5{\rm kpc}\times 50 \kms $   & 1.25 & (1)  \\
{\rm Sgr stream}	& $16{\rm kpc}\times 260 \kms $ & 20.8 & (2) \\
{\rm Magellanic stream} & $60{\rm kpc}\times 250 \kms $ & 75 & (3) \\
{\rm Ursa Minor} 	& $70 {\rm kpc}\times 200\kms $ & 70 & (4) \\
{\rm Fornax} 		& $138 {\rm kpc}\times 310\kms $& 213& (5) \\
{\rm M31 stream} 	& $150{\rm kpc}\times 20 \kms $ & 15 & (6)\\
{\rm Canis Major dSph}	& $15{\rm kpc}\times 200 \kms $ & 15 & (7)\\
\enddata
\tablerefs{(1) {Dinescu et al. 1999}; (2) {Ibata et al. 1997}; (3) 
{Kroupa \& Bastian 1997}; (4) {Schweizer et al. 1997} ; (5) 
{Piatek et al. 2002}; (6) {McConnachie et al. 2003}; (7) {Martin et al. 2004}.}
\end{deluxetable}

Suppose there was a population of hypothetical dwarf satellites 
in the outer halo with a specific angular momentum comparable to that of the orbits of LMC, Ursa Minor and Fornax.  
Suppose their initial mass $m_i$ is between Ursa Minor 
and the LMC ($10^{7-10}\msun$) at sometime $t_i$ between $4-14$Gyr. 
We integrate forward in time to answer the question where their remnants are.
Clearly the effect of orbital decay is maximized if we take the longest evolution time (10Gyr),
the highest initial satellite mass and smallest initial orbit $j_i=60{\rm kpc}\times 250\kms$.  
This is the case shown by the hatched region in Fig. 2 
with the satellite starting from the upper right corner and ending to the lower left with a mass $10^7\msun$.  
The vertical axis $S(t)={j(t)/200}$ is a characteristic orbital distance of the satellite,
expressed in terms of the specific angular momentum divided by a characteristic velocity $200\kms$.
This orbital distance is roughly the geometrical mean of the apocenter and pericenter.
Models are shown in order of increasing dynamical braking.
The upper shaded zone are models with between
exponential and linear massloss and a moderate evolution of the potential ($\infty \ge n \ge 1, p={1 \over 3}$), 
the lower shaded zone are models with between linear to accelerated massloss and a static potential 
($1>n>0.3,p=0$).  Qualitatively speaking the orbital decay appears to be only modest in all cases,
the remnants are generally not delivered to the inner galaxy.  

The condition for a satellite to deliver a low-mass substructure to the inner halo (cf. eq.15)
or a $10^6\msun$ black hole to the galaxy center 
is summarized in Fig.3.  The satellite
must be within 20kpc for the past Hubble time 
for a low-mass ($10^{7-9}\msun$) progenitor.  It could be at a modest distance
of 40-50kpc if the progenitor was very massive ($10^{10}\msun$) with a linear or accelerated massloss
($n<1$) and little evolution of the galactic potential ($p \ll 1$).
Progenitors of the inner halo substructures or central black hole cannot be 
on orbits of specific angular momentum comparable to the LMC, Ursa Minor or Fornax.
This illustrates the difficulty of making systems such as \omegacen as the nucleus of
a stripped-off dwarf galaxy starting from the very outer halo. 
Likewise it is difficult for a minor merger to bring in a million solar mass black hole
to the host galaxy center.  The progenitor's orbit must be radial and well-aimed 
at the galaxy center such that the progenitor's tangential velocity is  
$\le 1\kms$ if the satellite comes from an initial distance of 1Mpc.

\section{Comparison of various models for tides and satellite density profiles}

The basic feature of our dynamical friction models is that they 
bypass any information of the satellite internal density profile
and the varying tidal force on the satellite
by specifying the effective mass $\msat(t)$ as an explicit empirical function of time directly.  
This immediately brings up two questions, which we answer in the next two subsections: 
(1) Is our analytical model accurate enough?  Compared to a simulation with 
the more traditional "tidal peeling" approach, does our model reproduce the overall rate of the 
orbital decay for the same initial conditions?  
(2) Can our analytical model
be used to infer the underlying mass profile for satellites?  Is the inferred mass profile plausible
for observed satellites?  

\subsection{Inferring satellite rotation curves and comparing with observations}

Indeed we can infer the internal mass distribution, or internal circular velocity curve,
of the satellite using the tidal radius criteria.  More specificly
the internal circular motion $v_{\rm cir}$ at the tidal radius $r_t(t)$ should be 
in resonance with the satellite's orbital frequency at the pericenter $r_p(t)$, 
meaning that their angular frequencies or time scales are equal with 
\beq
{r_t \over v_{\rm cir} } = t_{cr}  = {r_p \over V_{\rm cir}  }, 
\eeq
where $t_{cr}$ is the crossing time, and the radii
\beq
r_t(t) = {G m(t) \over v_{\rm cir}^2}, \qquad r_p(t)= {j(t) \over u_p V_{\rm cir}},
\eeq
where the factor $u_p$
is the boosting factor of the pericenter velocity due to eccentricity.  If we define $\rho_t$ as the satellite's overall mean density at tidal radius $r_t$, 
and $\rho_{amb}$ and $M(t)$ as the average ambient density and the total mass
inside the pericenter $r_p$, then
\beq
m(t)={4\pi \rho_t r_t^3 \over 3}, 
M(t)={4\pi \rho_{amb} r_p^3 \over 3} ={V_{\rm cir}^2r_p \over G} , 
\eeq
Eliminate $r_p$ and $r_t$ with substitutions, we can rewrite the tidal criteria (eq. 23) as follows
\beq
\left({4\pi G\rho_t(t) \over 3}\right)^{-{1 \over 2}}
 =t_{cr} = \left({4\pi G\rho_{amb}(t) \over 3}\right)^{-{1 \over 2}}
\eeq
This means that as time progresses, the satellite mass $m(t)$
decreases (eq. 8), its orbital pericenter $r_p(t)$ decreases (eqs. 14, 24), 
the satellite's mean density $\rho_t(t)$ increases with the ambient
density $\rho_{amb}$ (eq. 26), 
hence the tidal radius $r_t(t)$ shrinks with the satellite's
mass $m(t)$, like what happens with peeling off an onion.  The tidal
peeling-off process effectively maps out $\rho_t$ as a function of $m$
implicitly through these equations.  The function $\rho_t(m)$ can
then be converted to the internal mass radial profile of the satellite
since $r_t(m) \propto \left[m/\rho_t(m)\right]^{1/3}$.  The circular
velocity as a function of enclosed mass $m$ can then be calculated
with $v_{\rm cir}^2(m) = G m/r_t(m)$.

The relation between $v_{\rm cir}$ and the mass $m(t)$ can also be shown more explicitly
as follows.  
From the tidal criteria (eq. 23) we have 
\beq
j(t) \propto r_p(t) V_{\rm cir}  
\propto V_{\rm cir}^{2} v_{\rm cir}^{-1} r_t 
\propto V_{\rm cir}^{2}v_{\rm cir}^{-3} m(t)
\eeq
since the tidal radius $r_t = G m v_{\rm cir}^{-2}$ (cf eq. 24).  
Substitute in the orbital evolution equation (eq. 6), and assume $\mu(t) \propto m(t)$, 
we have
\beq
{d \over dt} \left[m V_{\rm cir}^{2}v_{\rm cir}^{-3}\right] \propto
{dj(t) \over dt} \propto {m(t) V_{\rm cir}\over j(t)} \propto V_{\rm cir}^{-1}v_{\rm cir}^{3} .
\eeq
So there is a one-to-one relation between the massloss history $m(t)$ and the 
circular velocity at the tidal radius $v_{\rm cir}$ if we fix the host potential  
($V_{\rm cir}=cst$).  Interestingly this explains why
a satellite with flat rotation $v_{\rm cir} =cst$ leads to a linear 
mass loss history ${d\over dt}m(t)  = cst$.

A few circular velocity curves are shown in Fig.4 and Fig.5.
First we consider a hypothetical globular cluster of 
$10^6\msun$ on an orbit of apo-to-peri ratio of $15\kpc:3\kpc$,
which is within the inner galaxy although a slightly bigger orbit than \omegacen.
If this cluster is the end result of, say, lossing mass linearly with time 
from a $10^{10}\msun$ progenitor a Hubble time ago, then
the tidal criteria implies a mass profile of the progenitor,
shown by the circular velocity curve (labelled $n=1$) in Fig.4b.  
The progenitor must have a velocity curve close to that of an isothermal cored halo
with a tidal radius $r_t \sim 6.5\kpc$.  This implies an initial pericenter of 
about $r_p(t_i)={200\kms \over 80\kms}r_t \sim 16\kpc$ (cf. eq. 23).  So
the progenitor must start out on a small orbit with the first pericenter almost touching 
the inner halo of the host.  Such an orbit is a much smaller than the one that
the LMC and SMC are/were on, which has a pericenter about $40\kpc$ now and perhaps 
even further a Hubble time ago.  
This result holds qualitatively for a wide range of assumed massloss history. 
When we decrease $n$ from 30 to 0.3, the inferred initial tidal radius increases
from $r_t=4\kpc$ to $r_t=7.5\kpc$ (cf. Fig.4b), corresponding to a pericenter $r_p=8\kpc$ to
$r_p=20\kpc$.  The LMC has tidal radius of about $8-10\kpc$.  
So in order to deliver a final remnant to the inner 15 kpc,
we must start with a much smaller orbit and a much denser satellite than the LMC.

It is also remarkable that our predicted circular velocity curves
resemble very well 
those of observed dwarf galaxies, for a variety of initial and final
parameters of the satellite and the host galaxy.
Rotation curves are shown in Fig. 4a for several nearby dwarf galaxies
of virial mass $\sim 10^{10}\msun$:
DDO154 (Carnigan \& Purton 1998), NGC3109 (Jobin \& Carignan 1990), 
NGC5585 (Blais-Ouellette et al. 1999), NGC2976 (Simon et al. 2003).
Also indicated (by the locations of the small open circles)
are the core radius vs. the maximum rotation velocity for a sample of 
about 50 dwarf galaxies compiled in Sellwood (2000); only those with total mass
less than $10^{10}\msun$ are shown here.   

The fact that our models give reasonable mass profiles of satellites
justifies our empirical parametrization of the massloss history (eq. 14),
and gives a {\it physical meaning to our $n$ parameter}:  
$n=1$ resembles a cored isothermal model,
$n \gg 1$ gives an overall solid-body circular velocity curve, 
and $n \ll 1$ gives a Keplerian-like curve with a dense solid-body core. 

Observed dwarf galaxies typically have a solid-body core instead
of a cold dark matter cusp, with the density bounded typically between 
$0.25-0.0025\msun\pc^{-3}$; note the densest known dwarf galaxy Draco 
has a core density of $0.7\msun\pc^{-3}$ (Sellwood 2000).
These cores imply destruction of the typical dwarf galaxies once 
they decay to pericenters $r_p=3\kpc-30\kpc$ from
a Milky Way host.  
Among the observed dwarf galaxies, those with a core density 
comparable or lower than 
$0.01\msun\pc^{-3}$ (e.g., DDO154, NGC3109)
would be torn into tidal streams before reaching $15\kpc$ of 
the galaxy.

Similar calculations are also done for a Sgr-like remnant 
of $10^8\msun$ on an orbit of $40\kpc:8\kpc$ (Fig.5b). 
The results for the progenitor are shown and compared
with the observed dwarfs in a log-log plot (Fig.5a).  
Again the progenitors resemble the observed dwarfs, but again
the progenitor must originate from a small orbit.  

\subsection{Comparing the analytical model with "tidal peeling" simulations}

A key assumption of our model 
is that the effective mass is some simple empirical function of time (cf. eq. 14) with the normalization $\left<\beta\right>$ determined by the apo-to-perigalactic ratio (cf. eq. 21).
In reality the satellite's mass is determined by the tides at the pericenter,
and the dynamical friction force modulates periodically between pericenter and apocenter.
However, these short-time-scale variations are smoothed out
when averaged over a Hubble time
and almost do not contribute  to the evolution of the orbital angular momentum.  

To verify consistency of our empirical massloss history (eq. 14) with the tidal massloss history
we recompute the orbital decay using traditional tidal conditions, 
similar to Jiang \& Binney (2000) and Zhao (2002).
Here we fix $\ln \Lambda=2.5$ and show only two general cases with  
a satellite with a LMC-like flat rotation curve of amplitude $v_{\rm cir}=70\kms$
in a host halo of flat circular velocity curve of amplitude 
$V_{\rm cir}=200 \left({t \over 14}\right)^{1 \over 9}\kms$ (i.e. $p=1/9$). 
The satellite is launched on an eccentric orbit with initial specific 
angular momentum and apo-to-peri ratio being 
$j_i=45\kpc \times 280\kms$ and $C=3$ for model A,
and $j_i=20\kpc \times 350\kms$ and $C=8$ for model B, starting
from the pericenter (at $r_p=45\kpc$ or $20\kpc$); 
the pericenter radius and speed are determined by $C$ according to eq. (20),
which then determines the initial tidal radius of the satellite (cf eq. 23), 
and the mass (cf eq. 24).
The tidal criteria at the pericenters determines the evolution of the total mass of the satellite 
(cf. eq. 28), which should be slightly non-linear partly due to $V_{\rm cir} \propto t^{1/9}$
and partly due to circularization of the orbit.
The satellite orbit is followed for 10 Gyrs and shown in Fig.6.  
The orbit shrinks and becomes more circular than initially  due to dynamical friction and the satellite losses most of its initial mass.  
In Model A
it shrinks from 120kpc/45kpc to 15/10kpc, and the mass is reduced by a factor of about 5.
In Model B it shrinks 
from the initial apocenter/pericenter of 140kpc/20kpc to the final about 55kpc/15kpc and the mass is reduced by a factor of about 2.  

This is compared with our analytical model for the mass and angular momentum. 
The initial and final mass of the satellites $m_i$ and $m_0$ 
are taken from the above tidal-peeling simulations, and also for the 
the initial and final apo-to-peri ratios $C$, which 
we convert to the initial and final values for $\left<\beta\right>(C)$ with the help of eq. (21) or Fig.1b; generally $\left<\beta\right>_i \le \left<\beta\right>_0$ because of circularization;
together these specify the initial and final effective mass $\mu(t_i)$ and $\mu(t_0)$.
We then adopt a simple linear massloss $n=1$ law connecting the 
the initial and the final (effective) mass of the satellite.
The tidal criteria (eq. 28) suggests that 
such a linear massloss model would be rigorous 
for circular orbit in a static potential ($p=0$) since the satellite and 
the galaxy both have flat rotation curves.     
We then substitute this massloss history
in eq. (8) to predict the past angular momentum $j(t)$ from the present (smaller) value $j(t_0)=j_0$ backward.
\footnote{Predicting $j(t)$ from the initial (bigger) value 
$j(t_i)$ forward would be less stable or accurate once $j(t) \rightarrow 0$ because 
the relative error diverges.}   
The agreement of predictions from the two methods is clearly good  overall (cf. Fig.6b).  
The agreement is poorer for higher eccentricity orbit
with a stronger evolution in eccentricity. 
This is because 
our analytical method does not attempt to model the orbital circularization 
in any details apart from varying $\left<\beta\right>$.
 
We also check our models against previous fully self-consistent
live-halo simulations, i.e., simulations using live particles to represent
the halo of our galaxy.  we convert the initial conditions of these
simulations to calculate $S_i$ and $\mu_i$. This is shown in Fig.3.
The three simulations (Model A, F, K) of Jiang \& Binney (2000)
require a very massive ($10^{10-11}\msun$) progenitor of the Sgr far
away ($150-250\kpc$) from our Galaxy with an initial angular momentum
consistent with our linear massloss model ($n=1$, heavy dashed).  Here
we adopt initial apo-to-peri ratio $C \sim {200\kpc \over 60\kpc}$ and
$\left<\beta\right>_i=0.6$ from their simulations.  Jiang \& Binney
have also done semi-analytical modeling of orbital decay, and they
find models with $\ln\Lambda=8.5$ fit best.  The discrepancy with our
preferred value for $\ln\Lambda\sim 2.5$ are likely due to differences
in details of the analytical modeling: their Milky Way halo model has
an exponential truncation beyond 200 kpc, which reduces the halo's
density and dynamical friction at $150\kpc-250\kpc$ by about one
e-folding. Their defination of the tidal radius also implies a
systematicly stronger (up to a factor of two for Model A initially)
tide hence somewhat smaller satellite.  Also their semi-analytical
satellites loss mass somewhat faster than linear.

A rough agreement is also seen with self-consistent simulations
of \omegacen, where we adopt $C={60\kpc \over 1\kpc}$ hence  
$\left<\beta\right>_i=0.015$ for the best fit H4 model by Tsuchiya et al. (2003),
and $C={26\kpc \over 6\kpc}$ hence
$\left<\beta\right>_i=0.45$ for Bekki \& Freeman (2003).
There is some discrepancy with Bekki \& Freeman, perhaps due in part to our neglecting effects of the disk.  The final phase of their merger model
is perhaps too violent for ChandraSekhar's analytical description anyway.  
Indeed their orbit did not circularize, instead
the apo-to-peri ratio increased to $8\kpc:1\kpc$, 
somewhat larger than the observed orbit of \omegacen. 
Correcting down their orbital size 
would make better agreement with our prediction.

\section{Discussions}

\subsection{Effects of disk, bulge and orbital inclination}

One limitation of the current analysis is 
that we assume an isothermal dark matter plus stars model throughout the galaxy, hence
the dynamical friction effect of the disk and bulge are not modeled accurately.  However,  
the disk and bulge are not important for our conclusion because 
we predict mainly the orbital decay in the outer halo where 
$j>j_{\rm disk}=3000{\rm kpc~km/s}$.  Inside 15 kpc, our estimation of dynamical friction 
by an SIS model is inaccurate only for satellites on low inclination orbits.
If satellites come in random inclinations, it is more common to find high inclination orbits,
for which our models should be fairly accurate even inside 15 kpc.

\subsection{Effects of escaping stars}

In the part of our formulation where we derive the satellite mass profile,  
we assume a simplifying static picture that the satellite's mass is peeled off in successive layers at the shrinking tidal radius.   The picture in N-body simulations 
is more complicated, since satellite particles at all radii, e.g., the center of the satellite, 
could in principle be escaping at any time.  So eq. (26) sets only a lower limit on
the initial density of the satellite.  A more rigorous model should making this correction, e.g., by introducing an empirical factor to correct this as in Jiang \& Binney (2000). 

\subsection{Possible orbits of progenitors of the Sgr stream}

The Sgr dwarf and the Canis Major dwarf are the closest known dwarf galaxies, 
about 15 kpc from the center of the Milky Way and at the edge of the Milky Way disk.  
The Canis Major is on a (direct or retrograde) orbit 
slightly inclined from the plane of the Milky Way, and the Sgr is on a nearly polar orbit. 
Both orbits have a fairly low angular momentum with $S \sim 20\kpc$; the data on Sgr are 
more complete, and show that it oscillates between
10 kpc pericenter and 50 kpc apocenter.  Both contain 
several globular clusters.
It is possible that the two dwarfs are the stripped-down version of a more massive object,
which has dynamically decayed from the outer halo.  

Interestingly a possible 
extension of the Sgr has been reported recently in the SDSS data near the position of
the outer halo globular cluster NGC2419 (Newberg et al. 2003).
There is a stream-like enhancement of halo A-colored stars at the SDSS magnitude of $g_0=20.3$
in the plane of the Sgr's orbit, corresponding to a distance of 90 kpc.   
If this is true, it would imply that the Sgr
has changed its orbits in the past Hubble time.  There are two possible ways that
this could happen.  One is that the Sgr's orbit has been deflected by a massive
satellite, such as the LMC or SMC.  Indeed the orbits of the Sgr and the Magellanic Clouds
do overlap at the Galactic poles, and 
simple timing arguments show that these systems encounter or fly by
each other about 2.5 Gyrs ago at about 50 kpc on the North Galactic Pole if the rotation curve of the 
Milky Way is nearly flat (Zhao 1998).  The problem of this solution is that
it is rare for the Sgr to receive a strong enough deflection to bring down its orbit.

Another solution is that the Sgr has been a more massive system, which orbital decayed from
the outer halo (Jiang \& Binney 2000).  {\it Our Model B illustrates such an example of  
the progenitor of the Sgr} which had an initial mass comparable to the LMC ($10^{10}\msun$) and
was on an eccentric orbit with radius between $20-140\kpc$ (cf Fig.6).  This model is similar
to the Model K of Jiang \& Binney.
After a Hubble time the orbit decays to a small orbit very much like that of the Sgr
with peri-to-apo ratio of $10\kpc:50\kpc$.  Large amount of the material of the progenitor
is shed in the radius between $10-140\kpc$, the stream near NGC2419 at 90 kpc could be
part of this debris near one of the apocenters of the orbit.  Unfortunately the present 
model ends with a mass of $5\times 10^9\msun$, too large for the present-day Sgr.
Some fine tuning of initial conditions and detailed N-body simulations are clearly needed
to test this idea.

\subsection{Possible orbits of the progenitor of \omegacen}

We have mainly concentrated on the problem of getting rid of a satellite's angular momentum if it starts with a high angular momentum or orbital size $S_0\gg 15$kpc.
What would be the remnant distribution if a satellite is born with an initial orbital size $S_i<15$kpc?  
The stars in such a system are assembled in the inner halo 
from the start, e.g., by colliding an infalling gas cloud with the protogalactic gas clouds in the inner halo, 
(Fellhauer \& Kroupa 2002).  Or the stars form from extragalactic gas and descend on a very radial orbit,
penetrating the inner 15 kpc of the host halo from its very first pericentric passage.  

An intriguing example is \omegacen.
Unfortunately our analytical model is not suited for this system because it is presently 
on a low-inclination eccentric retrograde orbits between 1 and 6 kpc from the Galactic center (Dinescu et al. 1999), 
so the contribution of dynamical friction by the disk is important.  Also hydrodynamical friction
with the disk gas can play a role for an early-on partially gaseous satellite.  Nevertheless, 
if one applies simplisticly the tidal massloss and ChandraSekhar's dynamical friction in a spherical halo,
one finds that while it seems easy to peel off a satellite galaxy to make a central star cluster,
most simulations produce remnants on much larger orbits than \omegacen (Zhao 2002).  
It seems some fine tuning is required to select progenitors on very low angular momentum and/or low energy orbits:
the initial angular momentum needs to be low enough for 
the progenitor to penetrate into the inner halo or the present location of \omegacen on its 
very first pericentric passage.  This means that the initial orbital size $S_i^\omega$ of \omegacen is 
in between the present value of \omegacen $S_0 \sim 1.25$kpc and 
the boundary of the inner halo $R_{\rm disk}=15$kpc, or mathematically
\beq
1.25\kpc < S_i^\omega < 15\kpc.
\eeq

Most recently there have been several very encouraging attempts to model the dynamical and star formation history
of \omegacen by nearly self-consistent N-body simulations (Mizutani et al. 2003, Tsuchiya et al. 2003, Bekki \& Freeman 2003).
All are able to produce both a reasonable mass and orbit of the \omegacen after some trial and error 
with the initial parameters of the progenitor; 
many initial conditions lead to remnants, unlike \omegacen, beyond 10kpc of the Milky Way center. 
The favored initial orbit has a small orbital size $S_i$.
according to Tsuchiya et al. $j_i=60 {\rm kpc} \times 20 \kms =1200{\rm kpc}\kms$ (or $S_i=6$kpc) 
and  according to Bekki \& Freeman $j_i=25 {\rm kpc} \times 60 \kms =1500{\rm kpc}\kms$ (or $S_i=7.5$kpc).  
The small orbital size seems consistent with our expectation (cf. Fig.3).

Tsuchiya et al. launch satellites with various initial mass $(0.4-1.6)\times 10^{10}\msun$ 
and with either a King profile or a Hernquist profile from 60 kpc from the Milky Way center. 
They choose well-aimed nearly radial orbits, with an initial perigalactic radius about 1 kpc, 
much more radial than the present eccentric orbit.  
Massloss in their King model are similar to our exponential massloss models ($n=\infty$):
rapid in the beginning, and $\log(m)$ is roughly linear with time up to a mass of $10^8\msun$
when the satellite has too little mass to proceed with the orbital decay.
Massloss in the Hernquist model is closer to a $n=0.3$ model, linear in the beginning and 
rapid just before complete disruption (cf Fig.1a). 

Our comparison with Tsuchiya et al.'s numerical model would be fair apart from 
one theoretical concern.  The progenitor in their best simulation
is a two-component "nucleated" model with 
a rigid nucleus modeled by a extended-particle of $10^7\msun$ with a half mass radius of 35pc
on top of 
a live satellite of $0.8\times 10^{10}\msun$ with a Hernquist profile of half-mass radius $1.4$kpc;
the dynamical friction of the Hernquist halo helps to deliver the nucleus eventually to an orbit
similar to that of \omegacen.  However, a closer examination reveals a subtle inconsistency in making the nucleus rigid:  the tidal force from the Hernquist halo beats 
the self-gravity of this fluffy nucleus at its half-mass radius by a factor of a few, 
so it could not have stayed and being moved as one piece.  
Nevertheless, one could have used a more contact, hence more plausible model of 
the rigid nucleus, say, with a total mass of $3\times 10^6\msun$ and 
a smaller half-mass radius of 7pc, 
which are closer to the observed mass and half-mass radius of \omegacen.  
With this in mind, the formal inconsistency in Tsuchiya et al.'s best simulation seems to be
harmless, and their model shows that \omegacen could in principle be the remnant of a massive satellite on an orbit of initial apo-to-peri ratio of (about) 60kpc/2kpc.  

\section{Summary}

We have used a set of simple analytical models of dynamical friction and tidal massloss to 
explore the orbital decay for a dwarf satellite with a range of
initial specific angular momentum and massloss history.  These models greatly simplifies
the orbital dynamics and tidal interaction of satellites without lossing the accuracies
of more rigourous and sophisticated numerical simulations.
We follow the evolution of satellites in the mass-distance plane, and find 
generally very little evolution of specific angular momentum by dynamical friction.
The progenitors of inner halo globular clusters and substructures 
can not be born on orbits of comparable angular momentum 
as present-day halo satellite galaxies.  
The central cores of observed dwarfs are also not dense enough to survive the tides within 15 kpc. 
Any BHs in these satellites may also be difficult to reach and merger with the supermassive BH
in the host galaxy.  In general, satellite remnants
(BHs, globulars and streams) tend to hang up in the outer halo.

\acknowledgments 

I thank Ken Freeman, George Meylan, Jim Pringle, Floor van Leeuwen for enlightening discussions during the \omegacen conference at IoA,
and Oleg Gnedin, Mike Irwin, Pavel Kroupa, Mark Wilkinson for a careful reading of an earlier draft. I thank Vladimir Korchagin, Dana Dinescu \& Toshio Tsuchiya for patiently explaining their paper
to me, James Binney and Kenji Bekki for answering queries of their models.
This work is made possible by a special grant of research time from 
my 1-year-old MianMian and YiYi.

\begin{figure*}
\epsfxsize=8cm
\epsfbox{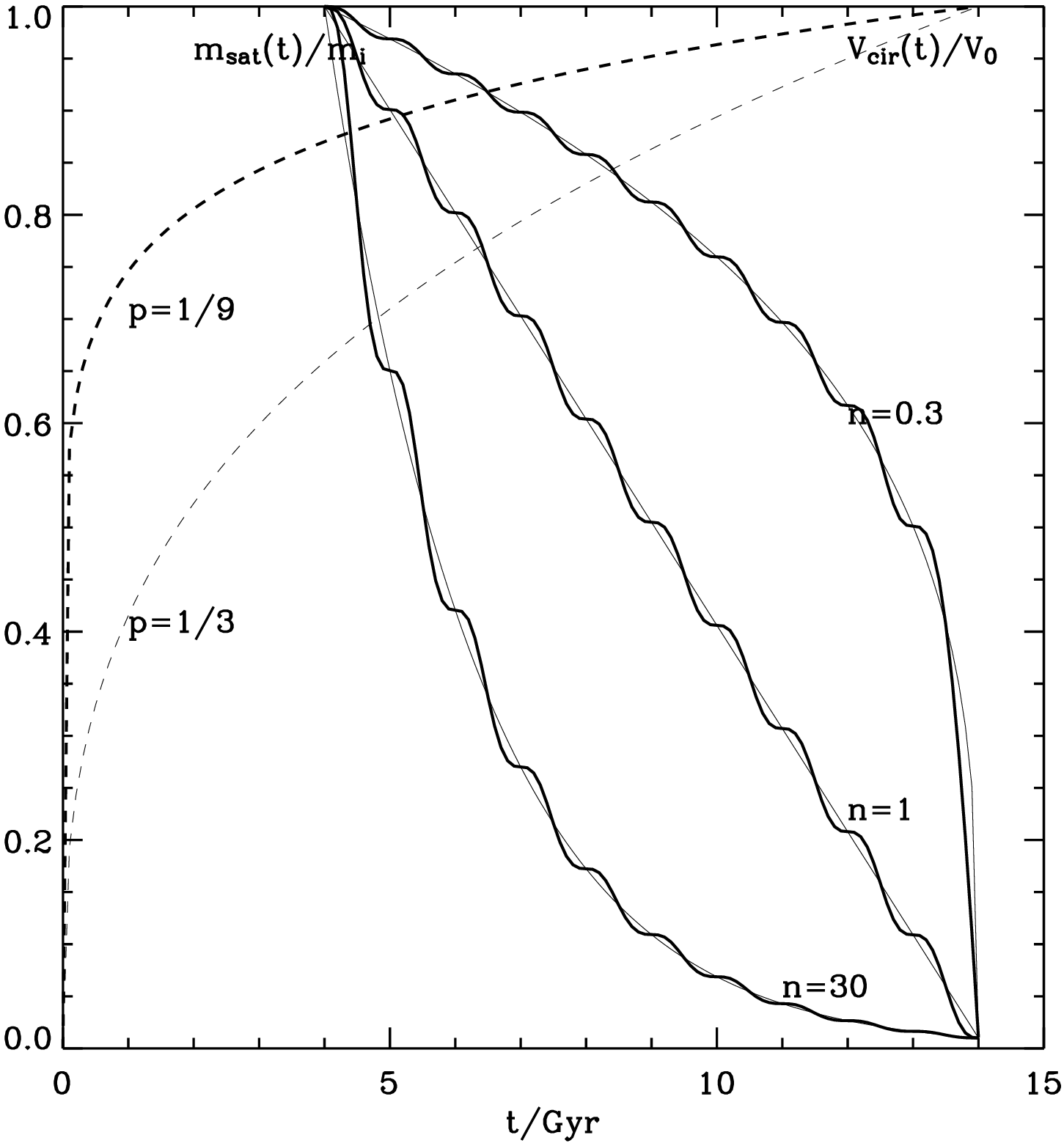}
\epsfxsize=8cm
\epsfbox{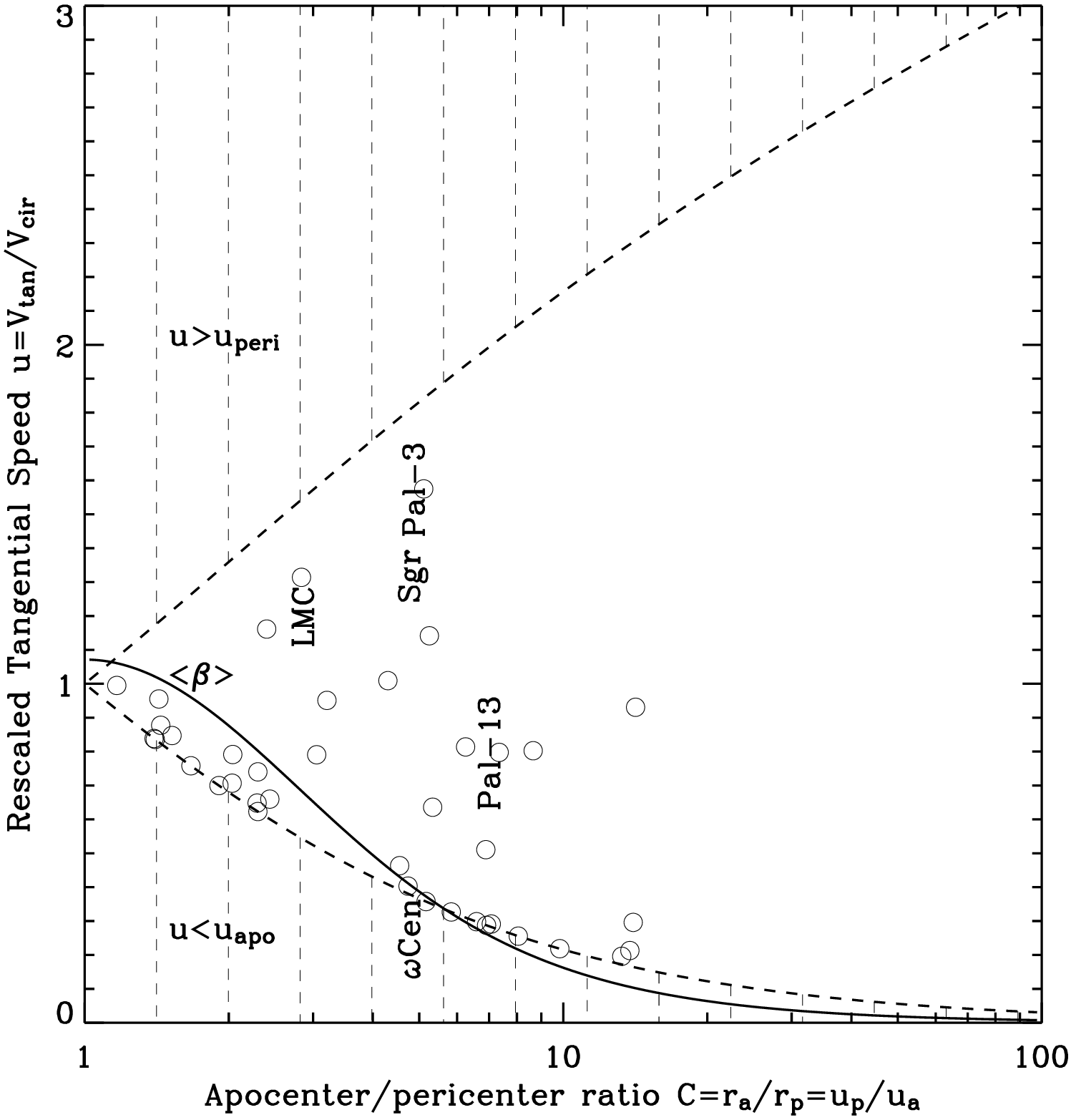}
\caption{
Panel {\bf (a)} 
shows the rescaled satellite mass $m(t)/m_i$ (thick solid) and effective mass $\mu(t)/\mu_i$ (thin solid) as a function of time
for three models (marked by their $n$-values),
and the rescaled host halo rotation speed $V_{\rm cir}(t)/V_0$ 
for the past Hubble time for two models (marked by their $p$-values).  
Panel {\bf (b)} shows the boundaries of
the rescaled tangential speed ${V_a \over V_{\rm cir}} \le u \le {V_p \over V_{\rm cir}}$,
(lower and upper dashed lines)
as functions of the apo-to-peri ratio $C={r_a \over r_p}={V_p \over V_a}$.
Also shown is the mean effective mass conversion factor $\left<\beta\right> = {\msat \over m}$ 
(solid line, cf. eq. 21), normalized to $\ln\Lambda=2.5$.  Note that the value of
$\left<\beta\right>$ varies {\it only} a factor of two among the Milky Way
satellites and globular clusters (circles); the pile-up of some clusters along  
the apocenter line is due to distance errors (data taken from Dinescu et al. 1999 and
references therein).
}
\end{figure*}

\begin{figure*}
\epsfxsize=14cm
\epsfbox{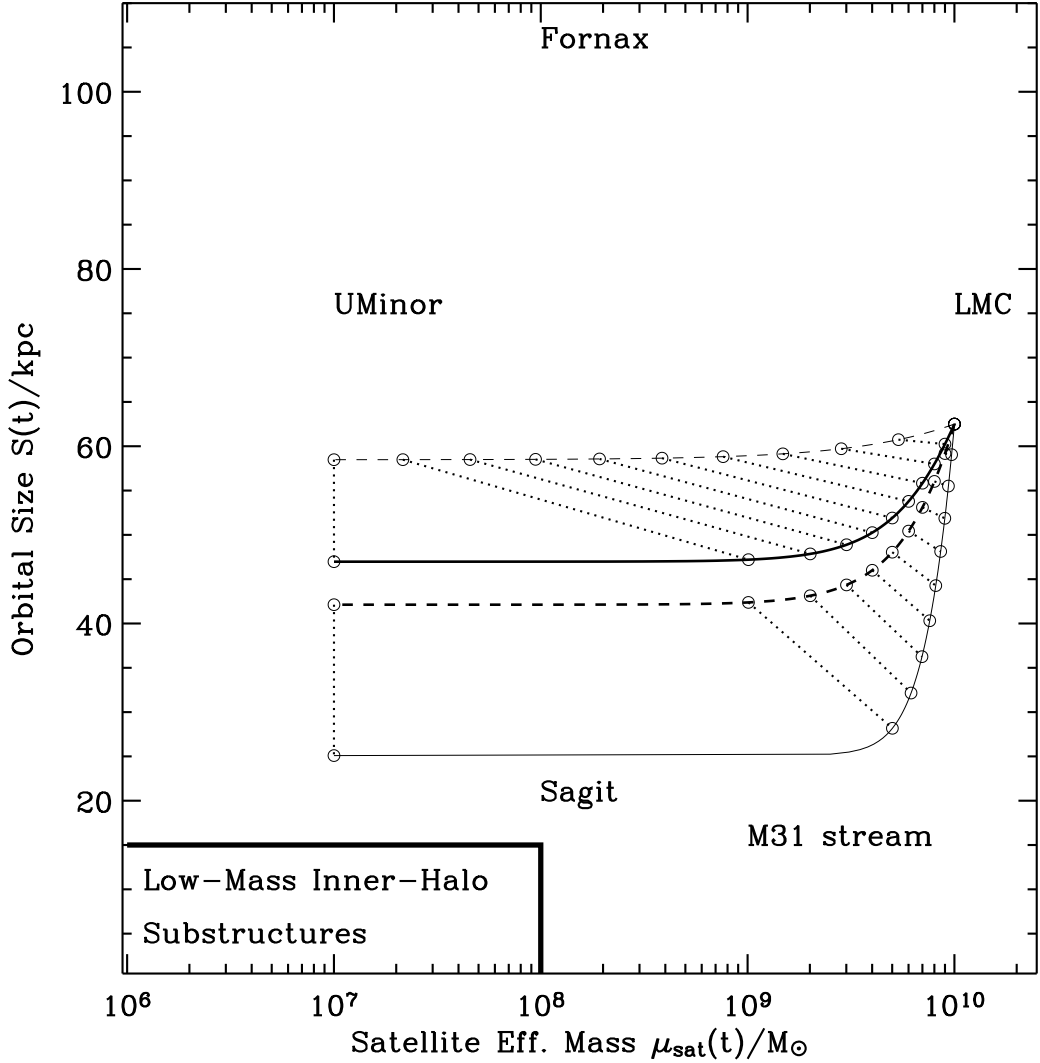}
\caption{
shows the predicted evolution histories of a satellite
in the plane of its effective mass $\msat(t)$ vs.  
characteristic orbital size $S(t) \equiv {j(t) \over 200\kms}$.
The hatched horn-like areas show models with moderate potential growth ($p={1\over 3}$, 
{\bf upper horn}) with a massloss between exponential ($n=\infty$, upper dashed boundary) and linear 
($n=1$, lower solid boundary), and models with static potential($p=0$, 
{\bf lower horn})
with a massloss between linear ($n=1$, upper dashed boundary) and accelerated 
($n=0.3$, lower solid boundary).
A massive satellite starts at $t_i=4$ Gyrs from the upper right corner with 
an effective mass $\mu(t_i)=\left<\beta\right>_im_i=10^{10}\msun$ 
with angular momentum $j_i=250\kms\times 50$kpc, and ends with an effective mass of $10^7\msun$.
For different assumptions of the massloss rate,  
the intermediate mass and position of the remnant are indicated
with a time step of 1 Gyr.  Note the failure to deliver remnants to the lower left corner.     
Also indicated are the estimated orbital size 
and the mass of the satellite galaxies of the Milky Way and M31.
}
\end{figure*}

\begin{figure*}
\epsfxsize=14cm
\epsfbox{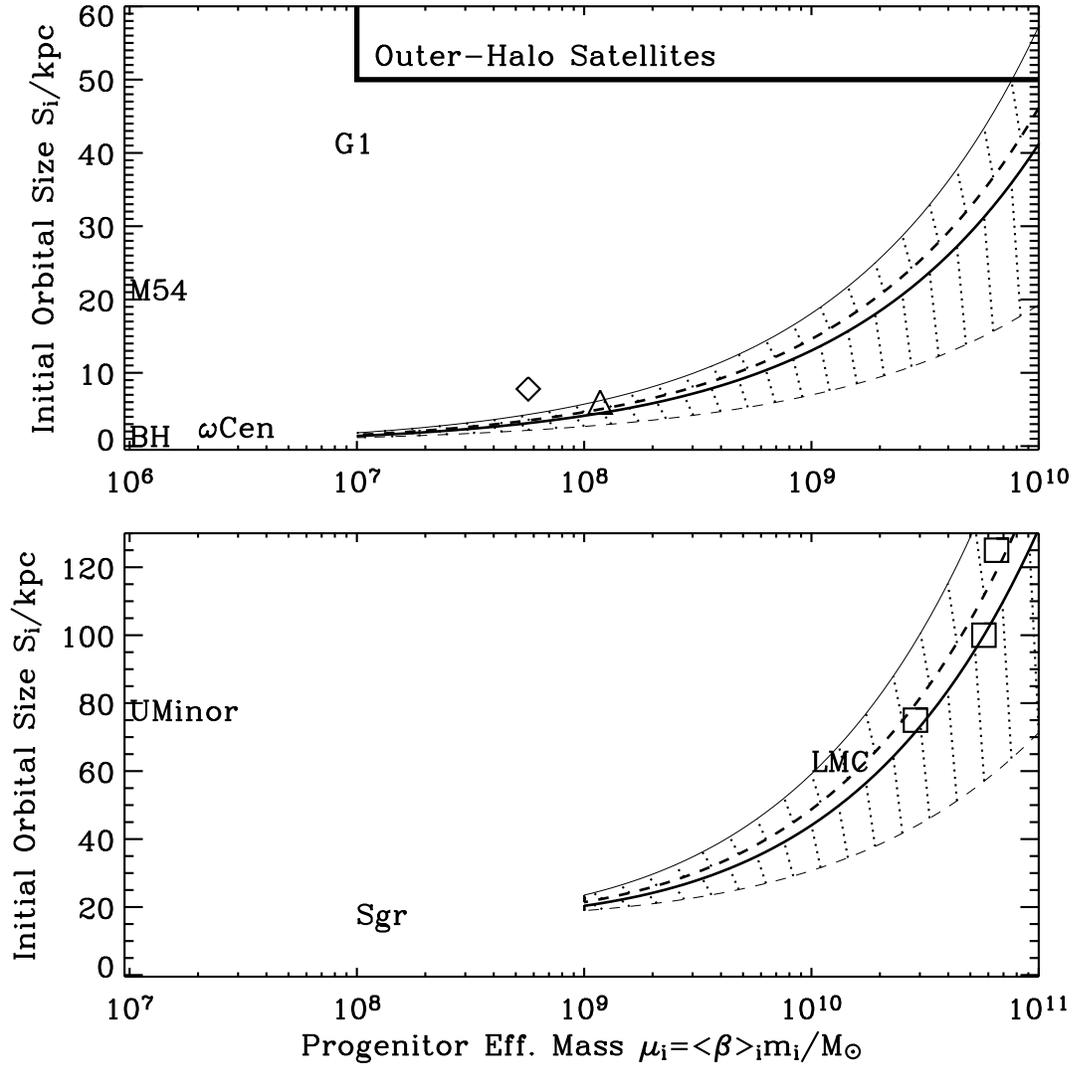}
\caption{
shows initial conditions 
to deliver a \omegacen-like $10^6\msun$ remnant to the Galactic center (upper panel) 
or deliver a Sgr-like remnant of $10^8\msun$ to the inner galaxy (lower panel). 
Dynamical friction works too slow for an object initially to the upper left of the shaded regions
in the plane of 
the initial effective mass of the progenitor $\mu_i$ vs. the initial orbital size $S_i=j_i/200$.
Different line types and shaded regions have the same meaning as in Figure 2 
(thick dashed line for a linear massloss in a static potential).
The symbols are simulations of Bekki \& Freeman (diamond) and Tsuchiya et al. (triangle) 
for the \omegacen, and Jiang \& Binney (squares for their Models A, F, K) for the Sgr; 
we assume the same Colomb logarithm $\ln\Lambda=2.5$ for both our model and these simulations.  
Also indicated are the present values for the several satellites
of the Milky Way and M31 and a central million solar mass BH.
}
\end{figure*}

\begin{figure*}
\epsfxsize=8cm
\epsfbox{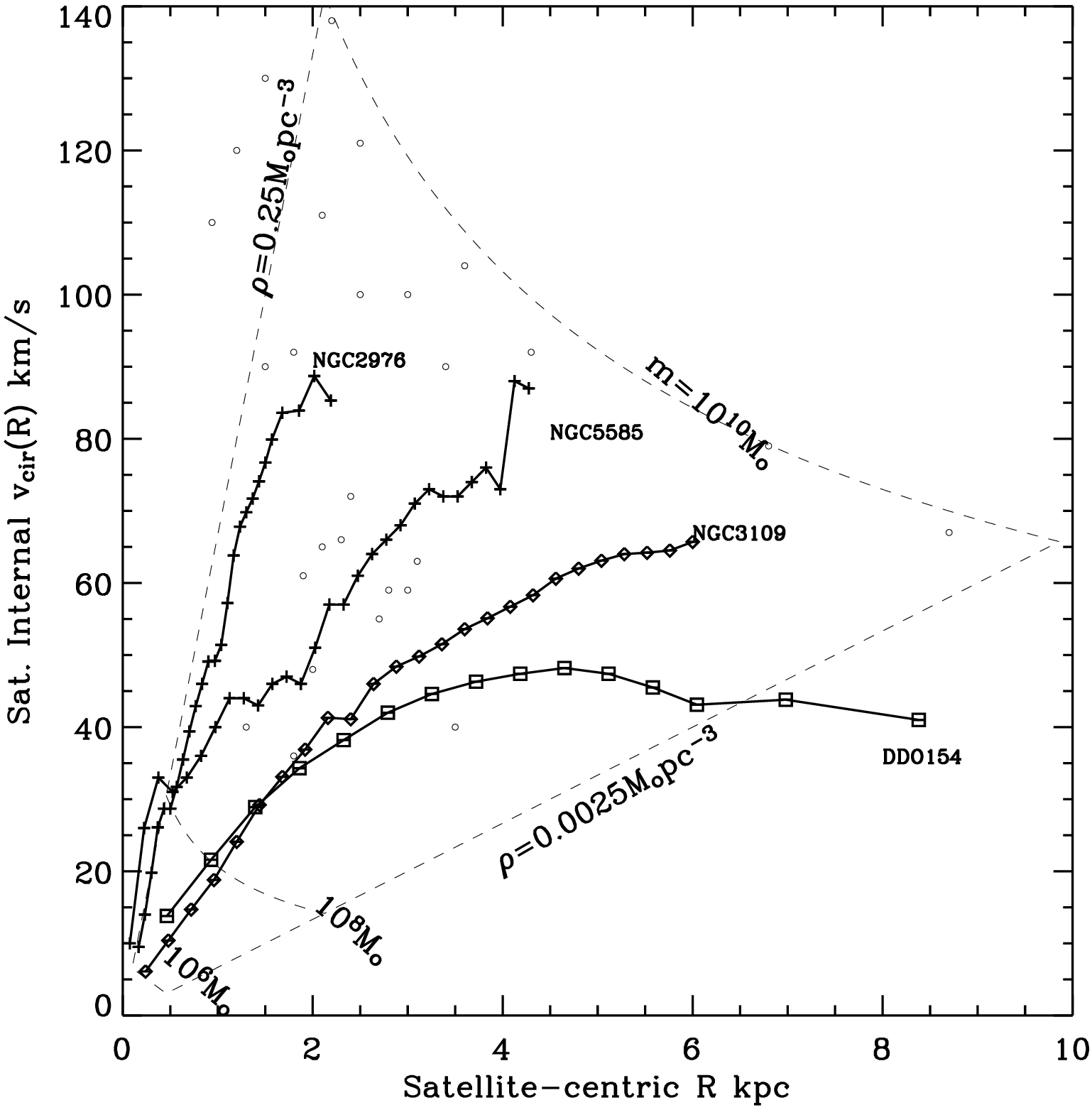}
\epsfxsize=8cm
\epsfbox{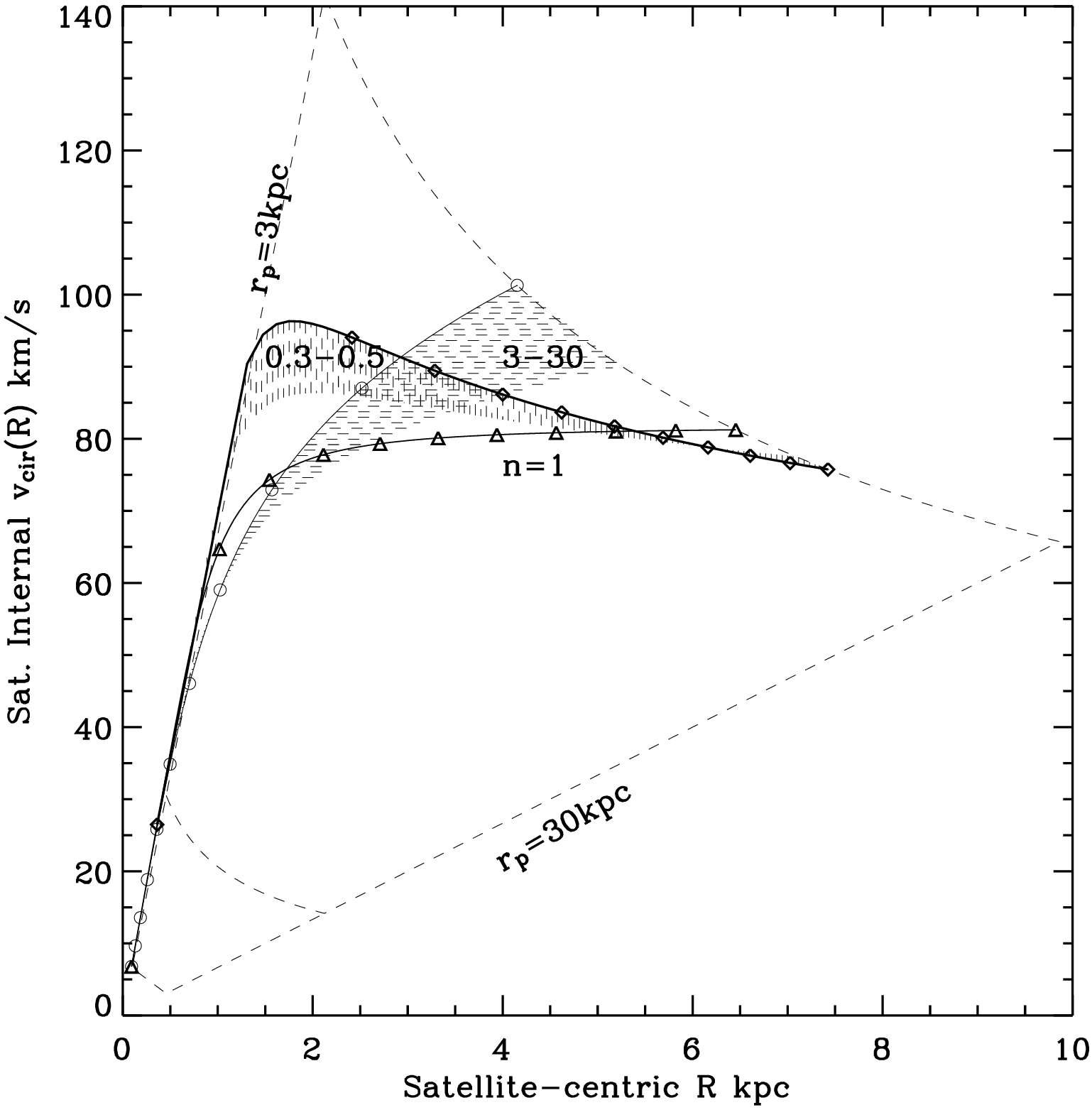}
\caption{({\bf panel a}) shows the rotation curves 
of four observed dwarf galaxies (as labelled), and
the core radius vs. maximum rotation velocity for 
another two dozen dwarf galaxies (small circles).
Also shown is a region bounded by solid body rotation
curves of a uniform volume density 
$\rho_t=0.25\msun\pc^{-3}$ or $0.0025\msun\pc^{-3}$ 
(two thin dashed lines from left to right), and Keplerian rotation
curves of a point mass of $10^{10}\msun$, $10^8\msun$, or $10^6\msun$ 
(three thin dashed curves from top to bottom).
({\bf panel b}) shows the inferred circular velocity curves of 
a $10^{10}\msun$ progenitor for different assumed massloss index $n$;
it is inferred so that a remnant of $10^6\msun$ 
is placed on an orbit with an apo-to-peri ratio of about $15\kpc:3\kpc$.
Symbols indicate the 
tidal radii at look-back time 1, 2, ..., 10 Gyrs for the $n=0.3$ model
(diamond), $n=1$ model (triangles), and $n=30$ model (circles).
Hatched regions are for $n=0.3-0.5$ and $n=3-30$.  
}
\end{figure*}

\begin{figure*}
\epsfxsize=8cm
\epsfbox{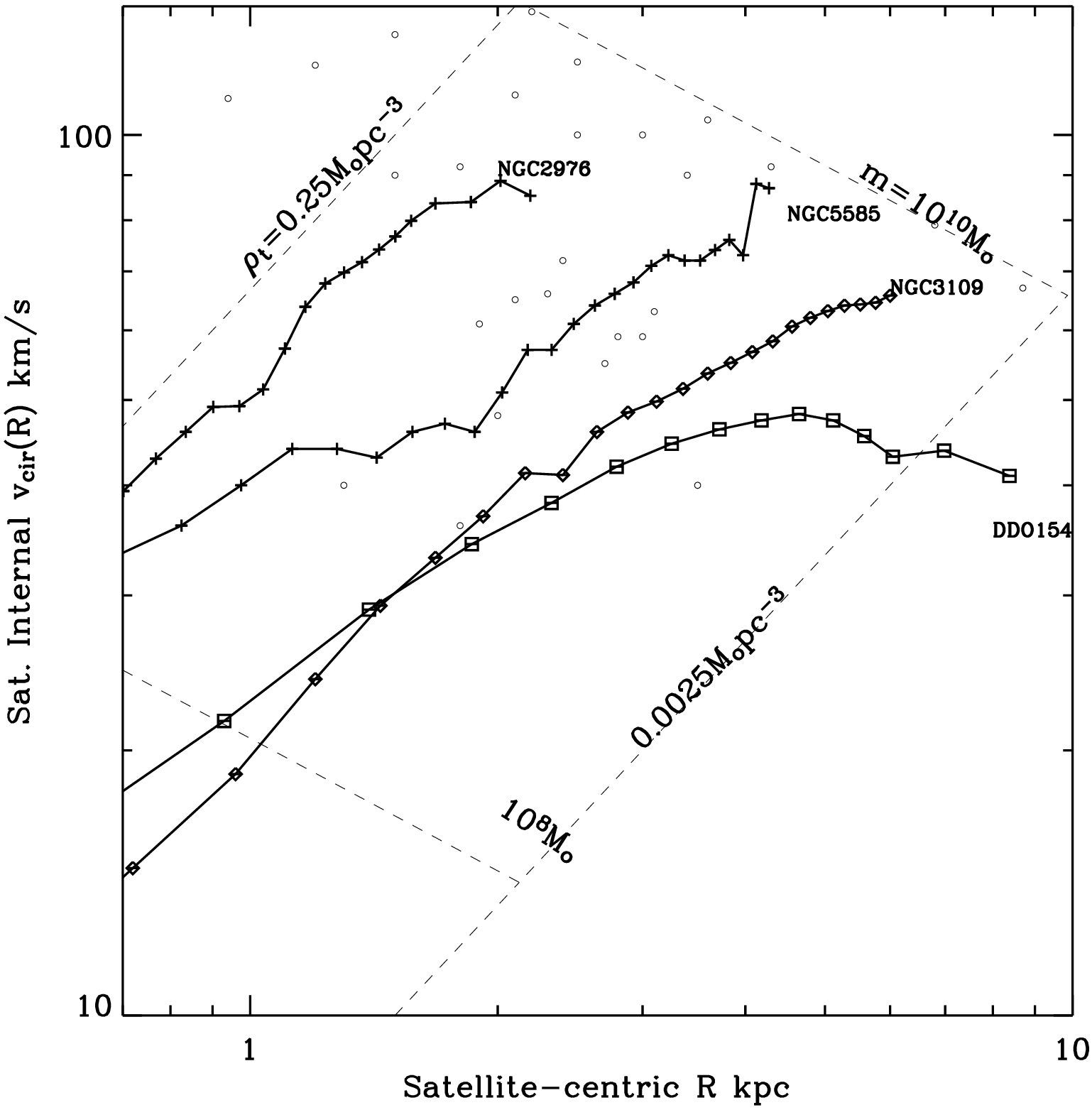}
\epsfxsize=8cm
\epsfbox{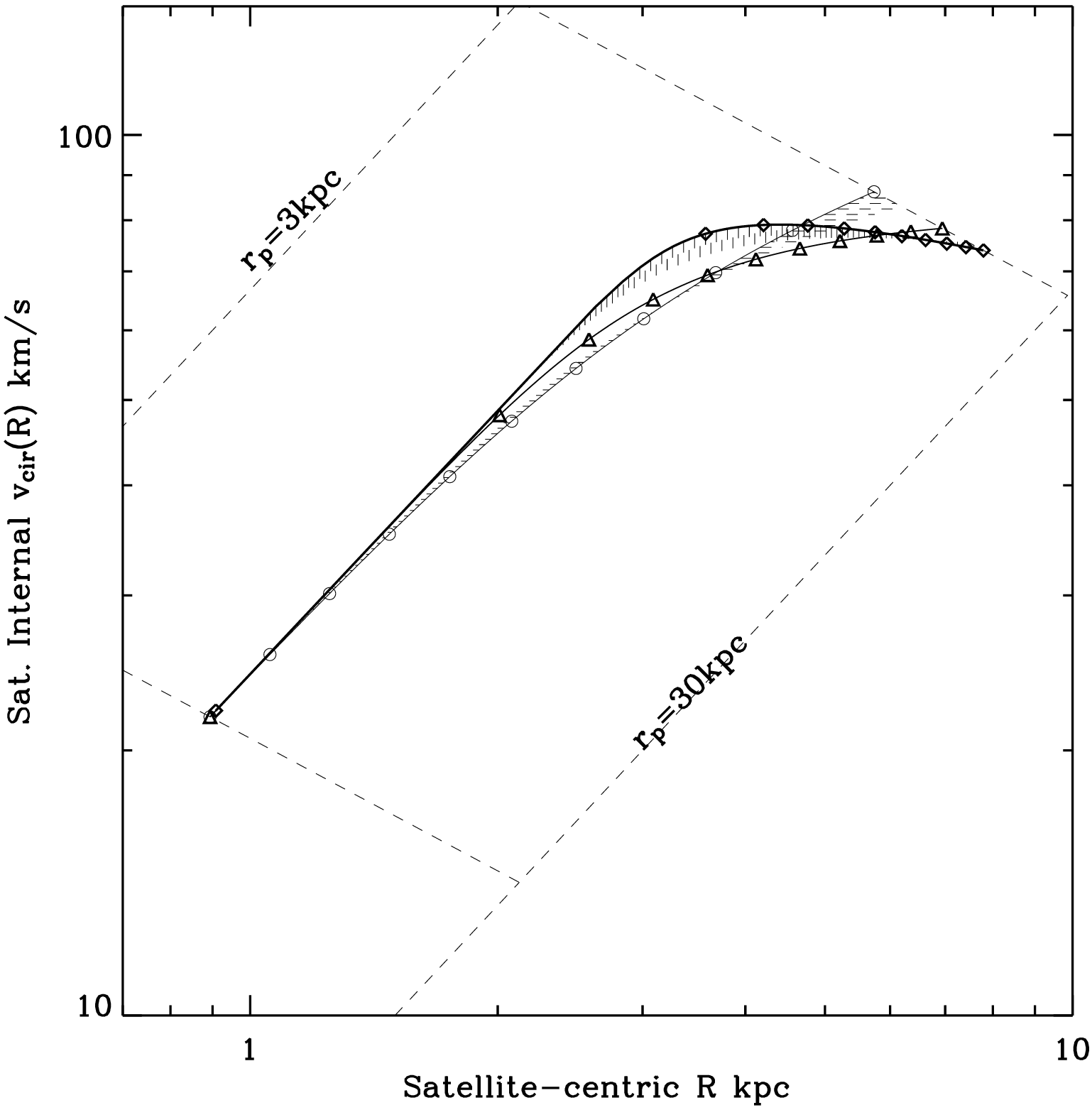}
\caption{Similar to previous figure, except in logarithmic scale, and for
({\bf panel b}) we assume the remnant is $10^8\msun$ 
on an orbit with $r_a:r_p=40\kpc:8\kpc$.
}
\end{figure*}

\begin{figure*}
\epsfxsize=8cm
\epsfbox{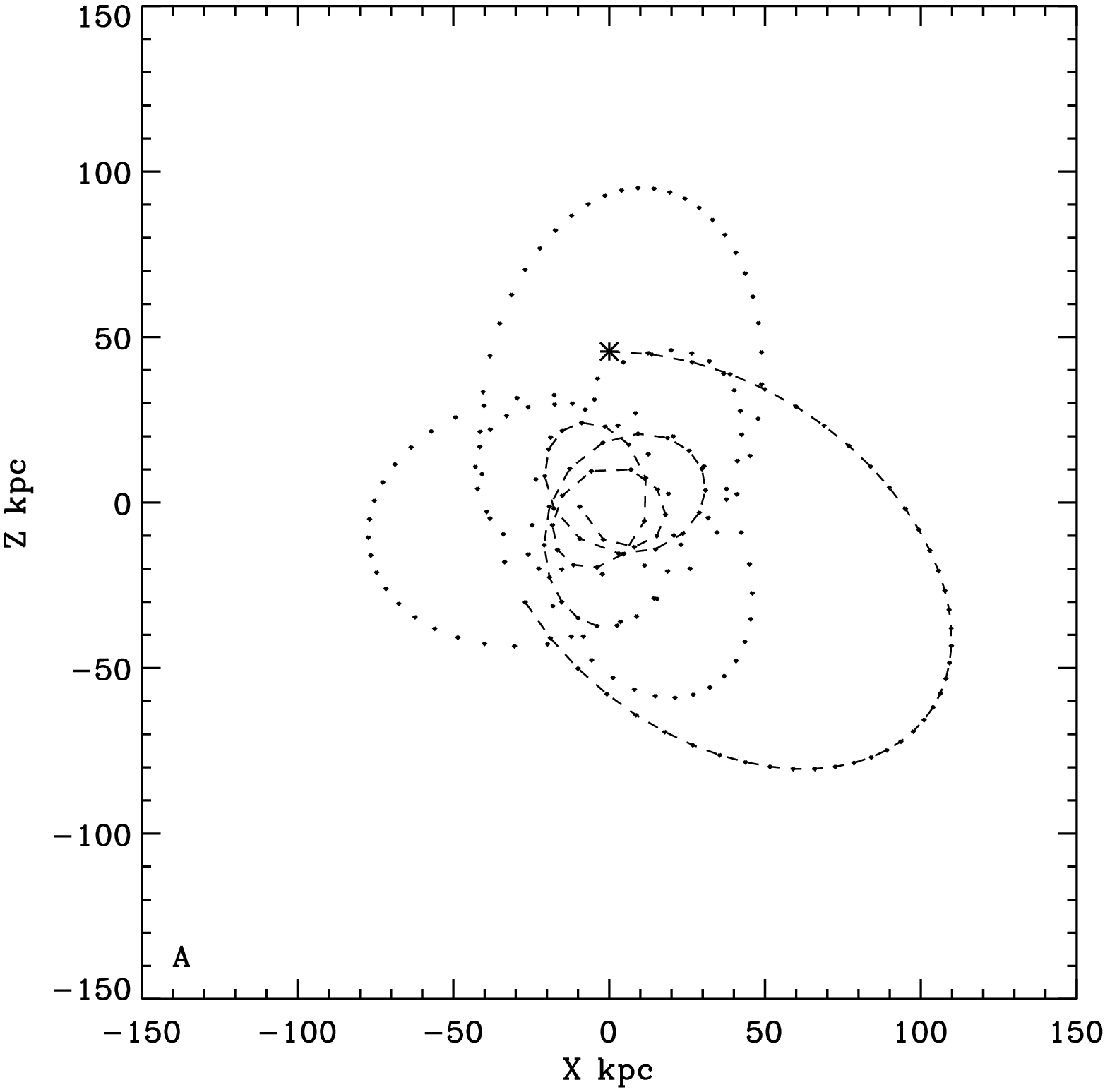}
\epsfxsize=8cm
\epsfbox{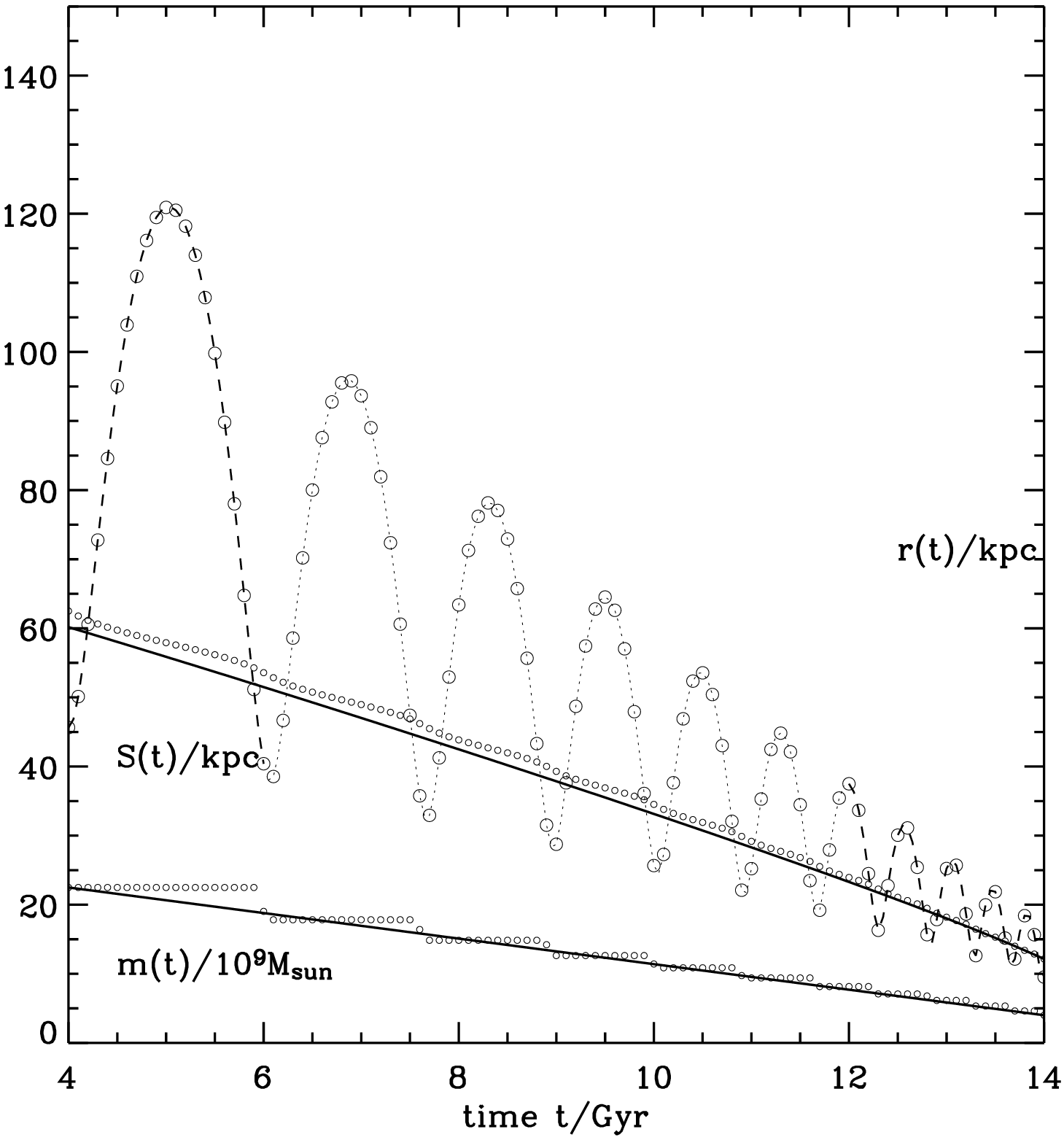}
\epsfxsize=8cm
\epsfbox{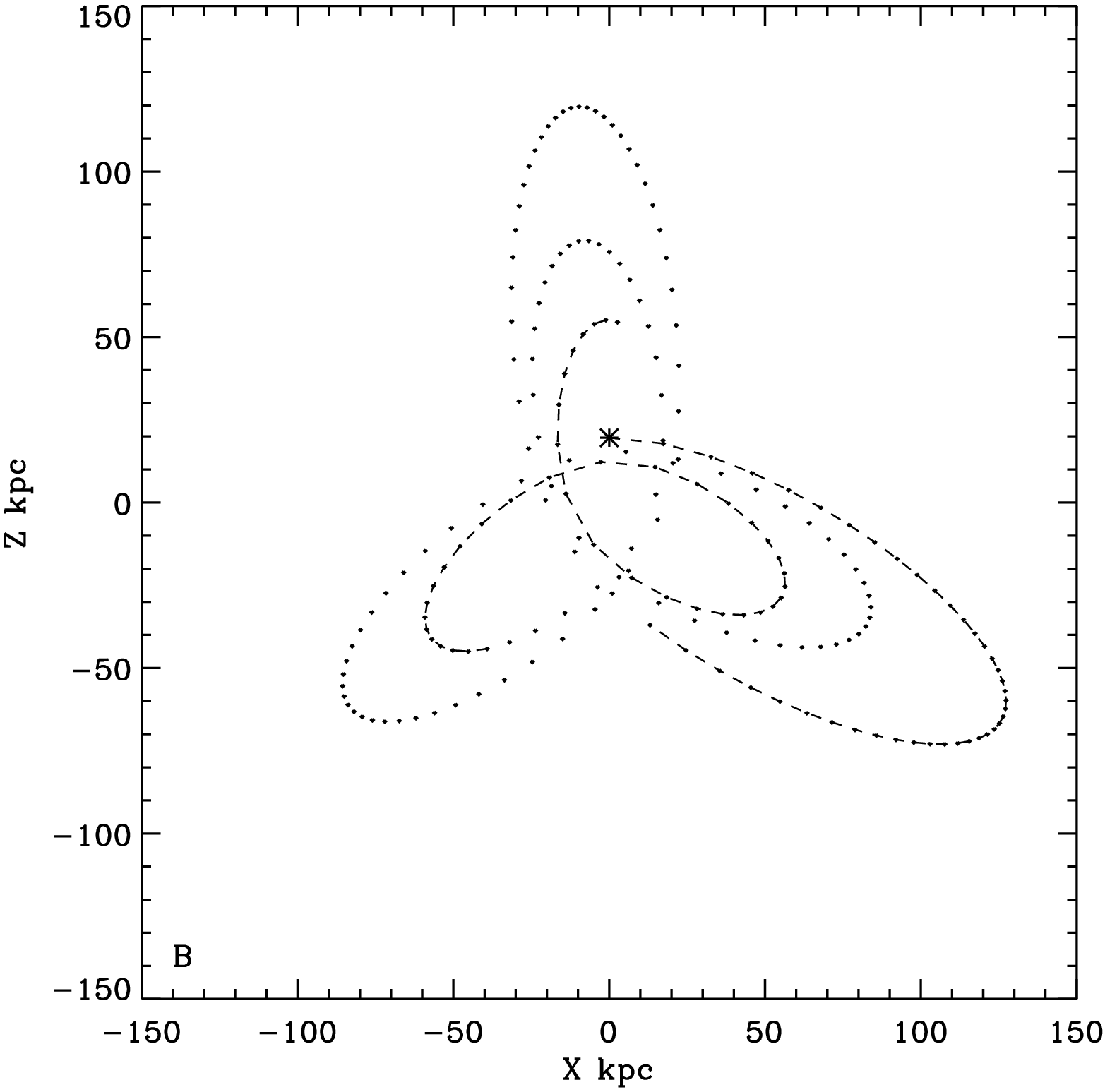}
\epsfxsize=8cm
\epsfbox{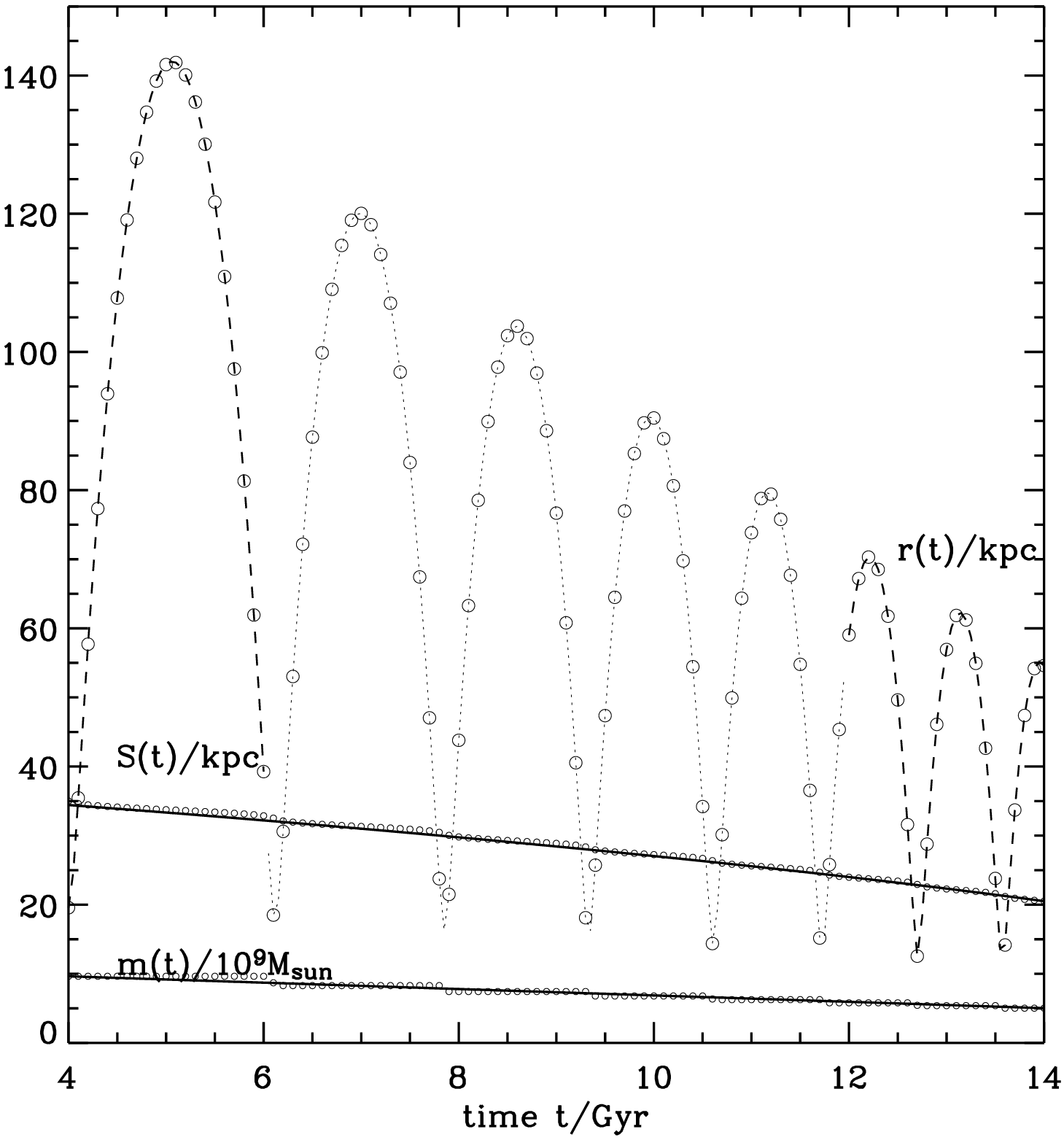}
\caption{shows an eccentric satellite orbit launched (from the positions marked by the star signs)
with initial apo-to-peri ratio $C=3$ (upper left) and $C=8$ (lower left, a model for the Sgr) 
in the X-Z plane from $t=4-6$Gyr (dashed) , $t=6-12$Gyr (dotted) and $t=12-14$Gyr (dashed).  
Also shown on the right are the evolution histories of the orbital radius $r(t)$, 
the orbital size $S(t)=j(t)/200\kms$ and the mass $m(t)$ by our analytical method (solid line) and 
the traditional method (points in even time step of 0.1Gyr).  
}
\end{figure*}

\vfill\eject

\end{document}